\documentclass[preprint,times,5p]{elsarticle}
\biboptions{sort&compress}
\newcommand{\vect}[1]{\mathbf{#1}}
\DeclareMathAlphabet\mathbfcal{OMS}{cmsy}{b}{n}
\usepackage{graphicx}  % needed for figures
\usepackage{dcolumn}   % needed for some tables
\usepackage{bm}        % for math
\usepackage{amssymb}   % for math
\usepackage{enumerate}
\usepackage{amsmath}
\usepackage{xfrac}
\usepackage{siunitx}
\usepackage{color}
\usepackage{placeins}
\usepackage{float}
\usepackage{caption}
\usepackage{subcaption}
\usepackage{hyperref}
\usepackage{natbib}

\begin{document}

\title{\textit{T}-matrix method for calculation of second-harmonic generation in clusters of spherical particles}

\date{\today}

\author[1]{Ivan Sekulic}
\ead{i.sekulic@ucl.ac.uk}

\author[1]{Jian Wei You}
\ead{j.you@ucl.ac.uk}

\author[1]{Nicolae C. Panoiu\corref{cor1}%
}
\ead{n.panoiu@ucl.ac.uk}

\cortext[cor1]{Corresponding author}

\address[1]{Department of Electronic and Electrical Engineering, University College London, Torrington Place, London WC1E 7JE, United Kingdom}

\begin{abstract}
In this article, we present a \textit{T}-matrix method for numerical computation of second-harmonic
generation from clusters of arbitrarily distributed spherical particles made of centrosymmetric
optical materials. The electromagnetic fields at the fundamental and second-harmonic (SH)
frequencies are expanded in series of vector spherical wave functions, and the single sphere
\textit{T}-matrix entries are computed by imposing field boundary conditions at the surface of the
particles. Different from previous approaches, we compute the SH fields by taking into account both
local surface and nonlocal bulk polarization sources, which allows one to accurately describe the
generation of SH in arbitrary clusters of spherical particles. Our numerical method can be used to
efficiently analyze clusters of spherical particles made of various optical materials, including
metallic, dielectric, semiconductor, and polaritonic materials.
\end{abstract}

\begin{keyword}\textit{T}-matrix method \sep plasmonics \sep second-harmonic generation \sep optical clusters \sep nanoparticles\end{keyword}
\maketitle

\section{Introduction}\label{INT}
Accurate description of light scattering from particles at the nanoscale has recently attracted
growing interest from the electromagnetics and optics communities, chiefly owing to a plethora of
applications made possible by our ability to control and manipulate light at this scale combined
with rapid advances in nanofabrication techniques. In this context, a major r\^{o}le is played by
localized surface plasmons, which are evanescent localized waves confined at the interface between
metallic nanoparticles and the surrounding dielectric medium \cite{Maier}. Indeed, at resonance,
the excitation of localized surface plasmons is accompanied by the generation of strongly enhanced
optical near-fields, a phenomenon that finds applications in many areas of science and engineering,
including surface-enhanced Raman spectroscopy, optical nanoantennas and sensors, optical
waveguides, metamaterials, and nonlinear optical microscopy
\cite{Cao1,Homola,Pendry,Liu,Ziolkowski,Kauranen,Butet,Panoiu,Boyd,Shen}.

Second-harmonic generation (SHG) is a second-order nonlinear optical process in which an incident
optical field oscillating at the fundamental frequency (FF), $\omega$, interacts with a nonlinear
medium and gives rise to a scattered optical field oscillating at the second-harmonic (SH)
frequency $\Omega=2\omega$ \cite{Bloembergen,Bozhevolnyi,Cao2,Cao3}. There are two principal
components of the nonlinear polarization, which are responsible for the SHG in so-called
centrosymmetric media, that is media that are invariant to inversion symmetry transformations. The
(local) surface nonlinear polarization is induced at the interface between the nanoparticle and the
surrounding environment, within a thin region containing just a few atomic layers, and where the
inversion symmetry is broken. The second component, the (nonlocal) bulk nonlinear polarization,
depends on the derivatives of the optical field components at the FF inside the nonlinear
scatterer. Second-harmonic radiation emitted from plasmonic nanoparticles is predominantly
generated by surface nonlinear polarization sources, but in the case of dielectric nanoparticles
surface and bulk nonlinear polarizations can have commensurate contributions \cite{Timbrell}.
Importantly, the resonant field enhancement due to excitation of localized surface plasmons in
metallic nanoparticles and Mie resonances in dielectric ones renders the light interaction with
nanoparticles an efficient source of SHG at the nanoscale.

It is therefore evident that developing numerical methods for efficiently finding accurate
full-wave solutions describing the light-matter interaction both at FF and SH frequencies is
particularly important. Analytical solutions to the SHG problem can be obtained only in a few
simple cases, such as a single sphere or an infinite cylinder. In a recent work \cite{Dadap1}, the
SHG theory for optically small centrosymmetric spherical particles has been presented, a theory
based on the so-called Rayleigh-Gans-Debye approximation, i.e., it is assumed that the field at the
FF is not perturbed during the scattering process. Nonlinear Mie theories regarding SHG and
sum-frequency generation (SFG) in a single sphere made of homogeneous centrosymmetric material have
also been developed \cite{Pavlyukh,Forestiere1,Beer1}. In the first of these studies the authors
take into account the surface nonlinear polarization source only, whereas in the last two works
both surface and bulk contributions of the nonlinear polarization are considered. In the pursuit of
the SHG/SFG solutions pertaining to arbitrarily shaped non-trivial particles and/or ensembles of
such particles, one has to resort to numerical methods.

Among various numerical techniques currently available in computational electromagnetism, such as
the finite-element method (FEM) \cite{Silvester,Jin} and finite-difference time-domain (FDTD)
method \cite{Yee,Taflove}, those exploiting the electromagnetic integral equations are particularly
effective since they produce comparatively smaller interaction matrices and inherently satisfy the
radiation boundary condition at infinity. The surface integral equations (SIEs) are usually
satisfied on the boundary-surface of the homogeneous particle
\cite{Mayergoyz,Hohenester1,Myroshnychenko}, by imposing appropriate boundary conditions, and
numerically solved in the framework of the method of moments (MoM) \cite{Harrington}. Another way
to solve SIEs is using the extended-boundary-condition (EBC) method, also called the null-field
approach, initially introduced for the analysis of perfect electric conductors \cite{Waterman1} and
subsequently extended to dielectrics \cite{Waterman2,Barber}, multiparticle systems
\cite{Peterson}, and efficient analysis of Raman scattering from molecules \cite{Schatz1}. In this
method, the SIEs are imposed on two surfaces defined inside and outside of the physical interface,
by exploiting the Huygens equivalence principle. In conjunction with vector spherical wave
functions (VSWFs) \cite{Varshalovich,Jackson} used in the series expansion of the fields, dyadic
Green's function, and unknown surface currents, EBC method leads to substantial computational
simplifications and reduction of memory requirements as compared to the MoM-SIE, especially in the
analysis of spherical particles.

The transfer matrix, relating the expansion coefficients of the incident and scattered fields, can
be easily derived using the EBC-VSWF approach. This method is also called the \textit{T}-matrix
method (TMM) or the multiple-scattering matrix (MSM) method \cite{Mishchenko0} (see also
Ref.~\cite{Mishchenko1} and the references therein). Note that the aforementioned numerical
techniques are less accurate when the size of the scatterer is of the order of a few nanometers or
less. In this case, quantum effects could become predominant, and classical numerical methods have
to be supplemented with quantum mechanical techniques \cite{Schatz2,Hohenester2,Deng,yp19jmmct}.

The MoM-SIE has been successfully implemented for the analysis of SHG from arbitrarily shaped
centrosymmetric homogeneous nanoparticles, by taking into consideration only the contribution of
the surface nonlinear polarization \cite{Makitalo} and the contribution of both surface and bulk
nonlinear polarizations \cite{Forestiere2}. These algorithms could be easily extended to the case
of multiparticle SHG, but even with state-of-the-art acceleration tools, such as multilevel fast
multipole algorithm \cite{Song} or adaptive cross approximation \cite{Bebendorf}, they are perhaps
prohibitively costly for very large number of particles. In recent studies \cite{Biris,BirisBis},
the TMM has been introduced in the frequency and time domains to calculate SHG from ensembles of
infinitely long nanowires made of centrosymmetric materials, with both the surface and bulk
components of the nonlinear polarization being included in the analysis. This work has been
extended to the SHG from spherical particles \cite{Xu}, but only the surface nonlinear polarization
has been considered.

In this paper, we present the \textit{T}-matrix method aimed at the calculation of SHG from a
cluster of spherical nanoparticles made of centrosymmetric materials. We build our method upon the
analytical work \cite{Forestiere2} and, differently from the approach introduced in \cite{Xu}, we
take into account both the local surface and the nonlocal bulk contributions to the nonlinear
polarization. We corroborate the theory with numerical examples and validate our method by
comparing the results with predictions obtained using a commercial software based on FEM. Our paper
is organized as follows. In the next section, we present the \textit{T}-matrix formalism both at
the FF and SH, then in Sec.~\ref{Examples} we illustrate on several examples the key features of
our method, and conclude this study with a summary of the main results.

\section{System geometry and \textit{T}-matrix method formulation of the problem}\label{MODmat}
In this section, we introduce the system configuration and its geometrical and material parameters,
together with the mathematical formalism of the \textit{T}-matrix method. The second-order
nonlinear scattering process under consideration is analyzed in two steps. In the first part, we
employ TMM to calculate the total field at the FF. We then use the computed FF field inside the
nanoparticles to determine the SH sources. With the nonlinear polarization sources at hand, the TMM
is employed again, and the SH field is computed. In this so-called undepleted-pump approximation of
the nonlinear scattering process we assume that there is no energy transfer from the SH back to the
FF. This is a correct assumption considering that the intensity of the SH field is several orders
of magnitude weaker than that of the fundamental field.
\begin{figure}[b!]
    \centering
    \includegraphics[width=\columnwidth]{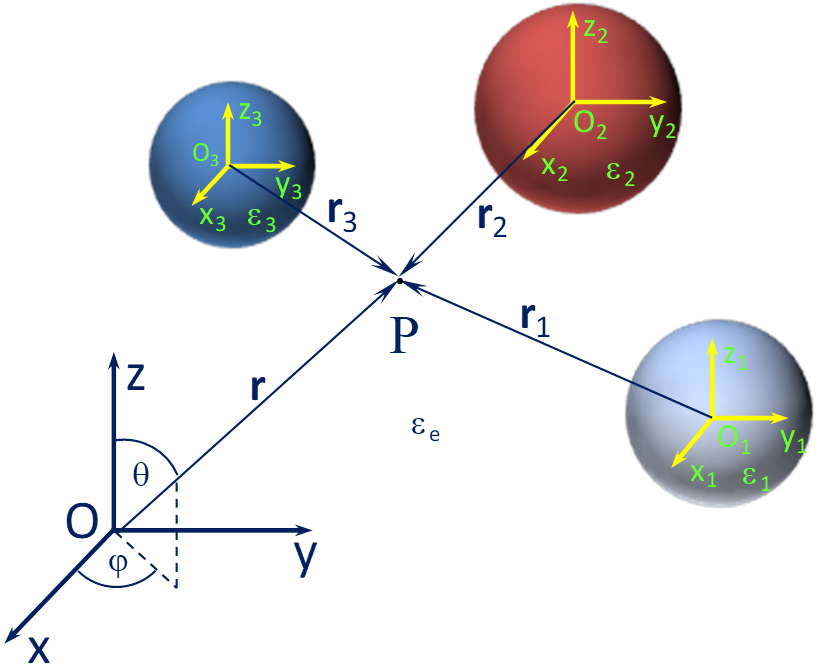}
\caption{Schematics of the configuration of a system of scatterers consisting of three nanospheres
embedded in a background medium. Also depicted are the global coordinate system with origin in $O$
and the local ones with origin in $O_{1}$, $O_{2}$, and $O_{3}$.}
    \label{fig:Spheres}
\end{figure}

\subsection{Geometrical configuration and material parameters}\label{Mod}
We seek to characterize the SHG from clusters of arbitrarily distributed spherical nanoparticles
made of homogeneous and isotropic centrosymmetric materials. The cluster is illuminated by an
incident monochromatic electromagnetic plane wave,
$\{\vect{E}_{inc}^{\omega},\vect{H}_{inc}^{\omega}\}$, with angular fundamental frequency $\omega$.
Also, we assume that all fields have a time-harmonic dependence, $e^{-i\omega t}$, which for the
sake of simplicity is suppressed throughout the manuscript. We index the set of particles with the
subscript $n$, which takes the values $n=1,2,\ldots,N$, with $N$ being the number of spherical
scatterers.

The $n^{th}$ sphere, with radius $R_n$, is made of (possible dispersive) material described by the
electric permittivity, $\epsilon_{n}$, and embedded in a lossless medium of permittivity
$\epsilon_e$. The magnetic permeabilities of the particles and the background medium are the same
and equal to the vacuum permeability, i.e. $\mu_{n} = \mu_e = \mu_0$. Furthermore, the nonlinear
polarization, $\vect{P}_n^{\Omega}$, responsible for the SHG, is governed by the third-rank
nonlinear susceptibility tensor, $\hat{\bm{\chi}}_n^{(2)}$. We label the origin of the global
coordinate system of the cluster with $O$, and to each particle we associate a local coordinate
system with origin $O_n$ located at the center of the corresponding sphere. The position of an
arbitrary point, $P$, can be expressed in two ways: either with respect to the global coordinate
system, \textit{via} the spherical coordinates $(r, \theta, \phi)$, or in the local coordinate
system associated to the $n^{th}$ sphere, using the spherical coordinates $(r_n, \theta_n, \phi_n)$
-- see Fig.~\ref{fig:Spheres}.

\subsection{Mathematical formulation of the \textit{T}-matrix method}\label{matfor}
In principle, many numerical algorithms for solving Maxwell equations can be used to derive the
transfer matrix relating the incident and scattered electromagnetic fields. If the particle under
consideration is a homogeneous sphere, the surface integral equations expressed in terms of VSWFs,
in conjunction with the EBC method, are an especially attractive choice. The decrease in the
requirements of computer memory, as compared to the MoM-SIE, comes from the fact that the system
matrix entries, which are related to certain integrals over the surface of the particle, can be
evaluated analytically. The entries of the \textit{T}-matrix computed in this way are equivalent to
the coefficients used in the Mie series expansion of the electromagnetic field.

\subsubsection{\textit{T}-matrix formalism at the fundamental frequency}\label{FF}
We divide the computational domain $V$ into an exterior subdomain $V_e$, occupied by the background
medium, and interior subdomains $V_{n}$, $n=1,2,\ldots,N$, occupied by the spherical particles. The
subdomains $V_e$ and $V_{n}$ are separated by oriented and closed boundaries of the nanoparticles,
$S_n$, defined with the unit normal vector $\hat{\vect{n}}_n$ pointing towards the exterior domain
$V_e$. The electromagnetic fields propagating at the FF satisfy the following Maxwell equations:
\begin{subequations}\label{eq:Maxwins}
\begin{align}
&\nabla \times \vect{H}_{n}^{\omega}(\vect{r})=-i\omega\epsilon_{n}^\omega \vect{E}_{n}^{\omega}(\vect{r}),\quad \vect{r} \in V_{n}\label{eq:Maxwins_a}, \\
&\nabla \times \vect{E}_{n}^{\omega}(\vect{r})=i\omega\mu_{0}\vect{H}_{n}^{\omega}(\vect{r}),\quad
\vect{r} \in V_{n}, \label{eq:Maxwins_b}
\end{align}
\end{subequations}
and
\begin{subequations}\label{eq:Maxwout}
\begin{align}
&\nabla \times \vect{H}_{s}^{\omega}(\vect{r})=-i\omega\epsilon_{e}\vect{E}_{s}^{\omega}(\vect{r}),\quad \vect{r}\in V_{e},\label{eq:Maxwout_a} \\
&\nabla \times \vect{E}_{s}^{\omega}(\vect{r})=i\omega\mu_{0}\vect{H}_{s}^{\omega}(\vect{r}), \quad
\vect{r} \in V_{e},\label{eq:Maxwout_b}
\end{align}
\end{subequations}
where $(\vect{E}_n^{\omega},\vect{H}_n^{\omega})$ are the fields inside the $n^{th}$ sphere and the
scattered electromagnetic fields, $(\vect{E}_s^{\omega},\vect{H}_s^{\omega})$, are equal to the
difference between the total exterior fields, $(\vect{E}_e^{\omega},\vect{H}_e^{\omega})$, and the
incident fields $(\vect{E}_{inc}^{\omega},\vect{H}_{inc}^{\omega})$:
\begin{subequations}\label{eq:ScatteredF}
\begin{align}
&\vect{H}_{s}^{\omega}(\vect{r})=\vect{H}_e^{\omega}(\vect{r}) - \vect{H}_{inc}^{\omega}(\vect{r}),\quad \vect{r}\in V_{e},\label{eq:ScatteredF_a} \\
&\vect{E}_{s}^{\omega}(\vect{r})=\vect{E}_e^{\omega}(\vect{r}) - \vect{E}_{inc}^{\omega}(\vect{r}),
\quad \vect{r}\in V_{e}.\label{eq:ScatteredF_b}
\end{align}
\end{subequations}

Equations \eqref{eq:Maxwins} and \eqref{eq:Maxwout} are supplemented with the following boundary
conditions expressing the continuity of the tangent components of the internal and external
electric and magnetic fields, $\hat{\vect{n}}_n
\times\vect{E}_{n}^{\omega}(\vect{r})\vert_{\mathbf{r}\in S_{n}}=\hat{\vect{n}}_n
\times\vect{E}_{e}^{\omega}(\vect{r})\vert_{\mathbf{r}\in S_{n}}$ and $\hat{\vect{n}}_n
\times\vect{H}_{n}^{\omega}(\vect{r})\vert_{\mathbf{r}\in S_{n}}=\hat{\vect{n}}_n
\times\vect{H}_{e}^{\omega}(\vect{r})\vert_{\mathbf{r}\in S_{n}}$, defined on the surfaces $S_n$
and written as:
\begin{subequations}\label{eq:Tancond}
\begin{align}
&\hat{\vect{n}}_n \times[\vect{E}_{n}^{\omega}(\vect{r})-\vect{E}_s^{\omega}(\vect{r})]=\hat{\vect{n}}_n \times \vect{E}_{inc}^{\omega}(\vect{r}),\quad \vect{r}  \in S_{n}\label{eq:Tancond_a}, \\
&\hat{\vect{n}}_n
\times[\vect{H}_{n}^{\omega}(\vect{r})-\vect{H}_s^{\omega}(\vect{r})]=\hat{\vect{n}}_n \times
\vect{H}_{inc}^{\omega}(\vect{r}), \quad \vect{r} \in S_{n}.\label{eq:Tancond_b}
\end{align}
\end{subequations}

In the framework of the \textit{T}-matrix method, we expand the incident, scattered, and internal
electric fields corresponding to each particle in terms of VSWFs, which represent a complete set of
vector functions:
\begin{subequations}\label{eq:ExpVSWF}
\begin{align}
&\vect{E}_{inc}^{\omega}(\vect{r}) = E_0 \sum\limits_{\nu\geq1} \Big[ q_{\nu}^{\omega}
\vect{M}_{\nu}^{(1)} (k_e^\omega \mathbf{r}) + p_{\nu}^{\omega} \vect{N}_{\nu}^{(1)} (k_e^\omega
\mathbf{r})\Big], \label{eq:IncidentVSWF} \\
&\vect{E}_{s}^{\omega}(\vect{r}) = E_0
\sum\limits_{n=1}^{N} \sum\limits_{\nu\geq1} \Big[ b_{n,\nu}^{\omega} \vect{M}_{\nu}^{(3)}
(k_e^\omega \mathbf{r}_n) + a_{n,\nu}^{\omega} \vect{N}_{\nu}^{(3)} (k_e^\omega \mathbf{r}_n)\Big], \label{eq:ScatteredVSWF} \\
&\vect{E}_{n}^{\omega}(\vect{r}) = E_0 \sum\limits_{\nu\geq1} \Big[ c_{n,\nu}^{\omega}
\vect{M}_{\nu}^{(1)} (k_{n}^\omega \mathbf{r}_n) + d_{n,\nu}^{\omega} \vect{N}_{\nu}^{(1)}
(k_{n}^\omega \mathbf{r}_n)\Big], \label{eq:InsideVSWF}
\end{align}
\end{subequations}
respectively, where $E_0$ is the amplitude of the incident wave, $k_e^\omega =\omega
\sqrt{\epsilon_e \mu_0}$ is the wave number of the background medium, and $k_{n}^\omega =\omega
\sqrt{\epsilon_{n}^\omega \mu_0}$ is the wave number of the medium in the particle $n$. The
parameters $\{q_{\nu}^{\omega},p_{\nu}^{\omega}\}$ represent the expansion coefficients in the
series expansion of the linearly polarized incident plane wave (see Appendix A for their
definition), whereas $\{b_{n,\nu}^{\omega},a_{n,\nu}^{\omega}\}$ and
$\{c_{n,\nu}^{\omega},d_{n,\nu}^{\omega}\}$ are the expansion coefficients of the scattered and
internal fields, respectively, associated with the $n^{th}$ spherical particle. We label the
spherical coordinates of the field-point $\vect{r}$, in the global coordinate system, with
$(r,\theta,\phi)$. The same point is defined in the local coordinate system associated to the
$n^{th}$ particle using the coordinates $(r_n,\theta_n,\phi_n)\equiv\mathbf{r}_n$ (see
Fig.~\ref{fig:Spheres}). The multi-index $\nu=(l,m)$ combines the orbital and azimuthal indices $l$
and $m$, respectively, so that the summation over this index is denoted as $\sum_{\nu\geq1} \equiv
\sum_{l\geq1}\sum_{m=-l}^{l}$.

The choice of VSWFs $\vect{M}_{\nu}^{(J)}$ and $\vect{N}_{\nu}^{(J)}$, $J=1,3$, in equations
\eqref{eq:ExpVSWF} is guided by the following considerations: the incident and the internal fields
are expanded in series of regular VSWFs $\vect{M}_{\nu}^{(1)}$ and $\vect{N}_{\nu}^{(1)}$, which
are finite at origin, whereas the scattered fields are expanded in series of outgoing VSWFs,
$\vect{M}_{\nu}^{(3)}$ and $\vect{N}_{\nu}^{(3)}$, which satisfy the radiation boundary condition
at infinity. The expansion of the magnetic fields in VSWFs is derived from equations
\eqref{eq:ExpVSWF} by using expressions \eqref{eq:Maxwins_b} and \eqref{eq:Maxwout_b}.

In the case of a single particle, the relation between the incident and the scattering coefficients
associated with the VSWF-expansion of the FF fields can be written as follows \cite{Mishchenko0}:
\begin{equation}
\mathbf{f}_n^{\omega}  = \mathbf{T}_n^\omega
\mathbf{g}_n^{\omega},
\label{eq:Tmatrixsingle}
\end{equation}
where $\mathbf{T}_n^\omega$ is the transfer matrix of the $n^{th}$ particle and has size of
$2N_{max} \times 2N_{max}$, and $\mathbf{f}_n^{\omega}=[b_{n,\nu}^\omega,
a_{n,\nu}^\omega]^{\mathrm{T}}$ and $\mathbf{g}_n^{\omega}=[q_{n,\nu}^\omega,
p_{n,\nu}^\omega]^{\mathrm{T}}$, $\nu = 1 \ldots N_{max}$, are vectors containing the expansion
coefficients of the scattered and incident fields expanded in the local coordinates, respectively.
Moreover, $N_{max}=l_{max}^{2}+2l_{max}$ is the number of Fourier coefficients in the series
expansion, with $l_{max}$ being the cut-off value for the orbital index, $l$.

For the particular case of a spherical particle, $\mathbf{T}_n^\omega$ is a diagonal matrix with
entries that can be calculated by imposing the boundary conditions \eqref{eq:Tancond} on the
VSWF-field expansions. These diagonal elements are the well-known transverse electric and
transverse magnetic Mie coefficients \cite{Mie}. If the particle is of arbitrary shape, the matrix
$\mathbf{T}_n^\omega$ is usually full.

The transfer matrix $\mathbf{T}_n^\omega$ pertaining to a scatterer of arbitrary shape can be
calculated using a matrix $\mathbf{Q}_n^{\omega(1,3)}$, which relates the incident and internal
field coefficients:
\begin{equation}
    \mathbf{h}_n^{\omega}  = -[\mathbf{Q}_n^{\omega(1,3)}]^{-1}
    \mathbf{g}_n^{\omega},
    \label{eq:Qmatrixsingle}
\end{equation}
where the vector $\mathbf{h}_n^{\omega}=[c_{n,\nu}^\omega, d_{n,\nu}^\omega]^{\mathrm{T}}$ contains
the internal field expansion coefficients. The relation \eqref{eq:Qmatrixsingle} is a direct
consequence of the equivalence principle applied to the region inside the particle. The notation
(1,3) indicates that the products of VSWFs in the integrals taken over the surface of the particle,
and which define the entries of the matrix $\mathbf{Q}_n^{\omega(1,3)}$, contain a regular and a
radiating function (for the definition of these integrals, see e.g.
Refs.~\cite{Forestiere3,Tsang}). In the case of axisymmetric particles, the 2D surface integrals
can be reduced to 1D line integrals, whereas in the particular case of a sphere, these integrations
can be evaluated analytically exploiting the orthogonality of VSWFs on a sphere. This reduced
computational complexity is perhaps the reason why the vast majority of the studies based on TMM
consider rotationally symmetric scatterers. It can be seen then that the transfer matrix
$\mathbf{T}_n^\omega$ for an arbitrarily shaped scatterer can be calculated using the following
relation:
\begin{equation}
     \mathbf{T}_n^\omega = -\mathbf{Q}_n^{\omega(1,1)}[\mathbf{Q}_n^{\omega(1,3)}]^{-1},
    \label{eq:TmatrixfromQ}
\end{equation}
where the integrals defining the matrix elements of $\mathbf{Q}_n^{\omega(1,1)}$ contain only
regular VSWFs.

It is well known that the standard TMM is less accurate when applied to situations in which the
near-field is highly inhomogeneous. This usually occurs when one deals with scatterers with sharp
edges and corners or objects with large permittivity. In order to circumvent this pitfall, a TMM
that exploits the \textit{discrete} VSWFs has been introduced \cite{Forestiere3}. However, if one
needs to determine very accurately the optical near-field in the vicinity of sharp tips, as
required for example in sensing applications, MoM-SIE is the method of choice \cite{Sekulic}.

In the case of a multiparticle system illuminated by a plane wave, the coupling between the
scatterers (multiple scattering) has to be taken into account. To this end, the total electric
field exciting the $n^{th}$ particle can be expressed as the sum of the incident field and the
field scattered from all the other particles:
\begin{align}\label{eq:IncFielnthPar}
\vect{E}_{n,ex}^{\omega}(\mathbf{r}) =& E_0\left[\vect{M}_{\nu}^{(1)}(k_e^\omega \mathbf{r}),\vect{N}_{\nu}^{(1)}(k_e^\omega \mathbf{r})\right]\mathbf{g}^{\omega}\\
&+E_0\sum\limits_{\substack{k=1\\k \ne n}}^{N}\left[\vect{M}_{\nu}^{(3)}(k_e^\omega
\mathbf{r}_k),\vect{N}_{\nu}^{(3)}(k_e^\omega \mathbf{r}_k)\right]\mathbf{f}_k^{\omega}, \quad
n=1,\ldots,N.\nonumber
\end{align}
This field is expanded around the origin of the $n^{th}$ sphere as:
\begin{align}\label{eq:IncFielnthParOrig}
\vect{E}_{n,ex}^{\omega}&(\mathbf{r}_n) = E_0\left[\vect{M}_{\nu}^{(1)}(k_e^\omega \mathbf{r}_n),\vect{N}_{\nu}^{(1)}(k_e^\omega \mathbf{r}_n)\right]\bm{\beta}_{n,0}^{\omega}\mathbf{g}^{\omega}\\
&+E_0\sum\limits_{\substack{k=1\\k \ne n}}^{N}\left[\vect{M}_{\nu}^{(1)}(k_e^\omega
\mathbf{r}_n),\vect{N}_{\nu}^{(1)}(k_e^\omega
\mathbf{r}_n)\right]\bm{\alpha}_{n,k}^{\omega}\mathbf{f}_k^{\omega}, \quad n=1,\ldots,N.\nonumber
\end{align}
In this equation, the matrices $\bm{\alpha}_{n,k}^{\omega}$ and $\bm{\beta}_{n,0}^{\omega}$, $n,k =
1, \ldots, N$, represent normalized irregular and regular translation-addition matrices,
respectively, evaluated at angular frequency $\omega$ \cite{Cruzan,Mackowski,Stout1}. The irregular
translation-addition matrix $\bm{\alpha}_{n,k}^{\omega}$ transforms the radiating (outgoing)
spherical harmonics $(J=3)$ expanded around the origin $O_k$, to the regular spherical harmonics
$(J=1)$ expanded around the origin $O_n$. The regular translation-addition matrix
$\bm{\beta}_{n,k}^{\omega}$ transforms the (regular or outgoing) VSWFs expressed in a coordinate
system with origin in $O_k$ to the same (regular or outgoing) VSWFs expressed in a coordinate
system with origin in $O_n$. As illustrated by \eqref{eq:IncFielnthParOrig}, for each scatterer
$n$, one uses the regular translation matrix $\bm{\beta}_{n,0}^{\omega}$ to translate the incident
field expansion around the origin of the global coordinate system, $O$, to the expansion around the
origin of the local coordinate system, $O_n$, associated to the $n^{th}$ spherical scatterer, i.e.
$\mathbf{g}_n^{\omega} = \bm{\beta}_{n,0}^{\omega} \mathbf{g}^{\omega}$.

The total field that excites the $n^{th}$ particle, $\vect{E}_{n,ex}^{\omega}(\vect{r}_n)$, can
itself be expanded into Fourier series of VSWFs, $\vect{M}_{\nu}^{(1)}(k_e^\omega \mathbf{r}_n)$
and $\vect{N}_{\nu}^{(1)}(k_e^\omega \mathbf{r}_n)$, with the corresponding Fourier coefficients
being $\mathbf{e}_n^{\omega}$. Then, as \eqref{eq:IncFielnthParOrig} indicates, these expansion
coefficients can be expressed as:
\begin{equation}\label{eq:IncFielnthParOrigCoeff}
\mathbf{e}_n^{\omega} = \bm{\beta}_{n,0}^{\omega}\mathbf{g}^{\omega} +\sum\limits_{\substack{k=1\\k
\ne n}}^{N} \bm{\alpha}_{n,k}^{\omega}\mathbf{f}_k^{\omega}, \quad n=1,\ldots,N.
\end{equation}
Multiplying to the left this equation with $\mathbf{T}_n^\omega$ and using the definition of the
transfer matrix for a single scatterer, one can recast \eqref{eq:IncFielnthParOrigCoeff} in terms
of the expansion coefficients of the scattering field as follows:
\begin{equation}\label{eq:ScattFielnthParOrigCoeff}
\mathbf{f}_n^{\omega} = \mathbf{T}_n^\omega \bm{\beta}_{n,0}^{\omega}\mathbf{g}^{\omega} +
\sum\limits_{\substack{k=1\\k \ne n}}^{N} \mathbf{T}_n^\omega
\bm{\alpha}_{n,k}^{\omega}\mathbf{f}_k^{\omega}, \quad n=1,\ldots,N.
\end{equation}

These equations can be collected together and expressed as a single matrix equation in the
following form \cite{Mishchenko0}:
\begin{equation}\label{eq:Tmatrixcluster}
\begin{bmatrix}
\textbf{I} & -\textbf{T}_{1}^\omega \bm{\alpha}_{1,2}^{\omega} & \cdots &  -\textbf{T}_{1}^\omega \bm{\alpha}_{1,N}^{\omega}\\
-\textbf{T}_{2}^\omega \bm{\alpha}_{2,1}^{\omega} & \textbf{I} & \cdots &  -\textbf{T}_{2}^\omega \bm{\alpha}_{2,N}^{\omega} \\
\vdots & \vdots & \ddots & \vdots \\
-\textbf{T}_{N}^\omega \bm{\alpha}_{N,1}^{\omega} & -\textbf{T}_{N}^\omega
\bm{\alpha}_{N,2}^{\omega} & \cdots & \textbf{I}
\end{bmatrix}
\begin{bmatrix}
\textbf{f}_{1}^\omega \\
\textbf{f}_{2}^\omega \\
\vdots \\
\textbf{f}_{N}^\omega
\end{bmatrix}
=
\begin{bmatrix}
\textbf{T}_{1}^\omega \bm{\beta}_{1,0}^{\omega} \textbf{g}^\omega \\
\textbf{T}_{2}^\omega \bm{\beta}_{2,0}^{\omega} \textbf{g}^\omega \\
\vdots \\
\textbf{T}_{N}^\omega \bm{\beta}_{N,0}^{\omega} \textbf{g}^\omega \\
\end{bmatrix},
\end{equation}
or in a more condensed form,
\begin{equation}
\mathbf{S}^{\omega} \mathbf{F}^{\omega}  = \mathbf{G}^{\omega}.
\label{eq:Scattmatclust}
\end{equation}
The matrix $\mathbf{S}^{\omega}$ is the scattering matrix of the cluster at the FF,
$\mathbf{F}^{\omega} = \{\mathbf{f}_n^\omega\}$, $n = 1 \ldots N$, is the vector containing the
unknowns, namely the expansion coefficients of the scattered field, and $\mathbf{G}^{\omega} =
\{\textbf{T}_{n}^\omega \bm{\beta}_{n,0}^{\omega} \textbf{g}^\omega\}$, $n = 1 \ldots N$, is the
vector containing the expansion coefficients associated with the incident field. The unknown
scattering field coefficients corresponding to each particle are obtained by simply solving the
linear system \eqref{eq:Tmatrixcluster}.

The scattering cross-section (SCS) and absorption cross-section (ACS), calculated at frequency
$\omega$, are then easily obtained from these scattering field coefficients using relatively simple
relations \cite{Stout1,Stout2}:
\begin{equation}
\sigma_{sca}^{\omega} = \frac{1}{(k_e^\omega)^2}\sum\limits_{n,k=1}^{N}
\mathfrak{Re}\{\textbf{f}_{n}^{\omega*}\bm{\beta}_{n,k}^{\omega}\textbf{f}_{k}^{\omega}\},
\label{eq:ScattcrosssectFF}
\end{equation}
\begin{equation}
\sigma_{abs}^{\omega} = \frac{1}{(k_e^\omega)^2}\sum\limits_{n=1}^{N}
\textbf{f}_{n}^{\omega*}\bm{\Gamma}_n^{\omega}\textbf{f}_{n}^{\omega}, \label{eq:AbscrosssectFF}
\end{equation}
where the symbol ``$*$'' denotes complex conjugation, $\mathfrak{Re}\{z\}$ is the real part of the
complex number $z$, and $\bm{\Gamma}_n^{\omega}$ is a matrix that transforms the scattering
coefficients to the internal field expansion coefficients. It can be easily calculated from the
field boundary conditions (see, e.g. Ref.~\cite{Stout1,Stout2}).

\subsubsection{\textit{T}-matrix formalism at the second-harmonic frequency}\label{SH}
In the second step of our numerical method, we use the fields at the FF, computed at the linear
step just described, and subsequently compute the sources at the SH frequency. Here, we focus on
SHG analysis from centrosymmetric materials widely used in nonlinear optics, including noble metals
(Au, Ag) and dielectrics (Si, $\mathrm{SiO_{2}}$) in which the crystal lattice is invariant upon
inversion symmetry transformations. Due to the centrosymmetry of the material, the (local) bulk
second-order susceptibility tensor $\hat{\bm{\chi}}^{(2)}$ identically vanishes inside the
particle. However, the inversion symmetry property is locally broken at the surface of the
centrosymmetric material \cite{Kauranen,Butet,Panoiu,Boyd,Shen}, and the corresponding surface SHG
is described by a third-rank nonlinear susceptibility tensor, $\hat{\boldsymbol{\chi}}_s^{(2)}$. In
the bulk of the centrosymmetric material, the (nonlocal) sources of SH consist of nonlinear
electric quadrupoles and nonlinear magnetic dipoles, the corresponding SHG being described by a
fourth-rank nonlinear susceptibility tensor $\hat{\boldsymbol{\chi}}_b^{(2)}$.

Putting these ideas together, the nonlinear polarization, $\vect{P}^{\Omega}(\mathbf{r})$, that
generates the SH field can be represented as the sum of the local surface and nonlocal bulk
contributions as:
\begin{align}\label{eq:SHPolarization}
\vect{P}^{\Omega}(\mathbf{r}) =& \vect{P}_s^{\Omega}(\mathbf{r}) + \vect{P}_b^{\Omega}(\mathbf{r})
=\epsilon_0 \hat{\boldsymbol{\chi}}_s^{(2)}:\vect{E}^{\omega}(\mathbf{r})
\vect{E}^{\omega}(\mathbf{r})\delta(\mathbf{r}-\mathbf{r}_{s}) \nonumber \\ & + \epsilon_0
\hat{\boldsymbol{\chi}}_b^{(2)} \vdots \vect{E}^{\omega}(\mathbf{r})
[\nabla\vect{E}^{\omega}(\mathbf{r})]\left.\right|_{\mathbf{r}\in V},
\end{align}
where $S=\bigcup_{n=1}^{N}S_{n}$ ($V=\bigcup_{n=1}^{N}V_{n}$) is the total surface (volume) of the
particles and $\mathbf{r}_{s}\in S$ defines the surface $S$. In equation \eqref{eq:SHPolarization},
the factors $\vect{E}^{\omega}(\mathbf{r})\vect{E}^{\omega}(\mathbf{r})$ and
$\vect{E}^{\omega}(\mathbf{r})[\nabla\vect{E}^{\omega}(\mathbf{r})]$ represent second- and
third-rank tensors, respectively, so that the nonlinear polarization in \eqref{eq:SHPolarization}
can be written componentwise as:
\begin{equation}\label{eq:SHPolarizationcomp}
    P_i^{\Omega} =\epsilon_0 \hat{\chi}_{ijk}^{(2)} E_j^{\omega}
    E_k^{\omega}\delta(\mathbf{r}-\mathbf{r}_{s}) + \epsilon_0
    \hat{\chi}_{ijkl}^{(2)} E_j^{\omega}
    \nabla_k E_l^{\omega}\left.\right|_{\mathbf{r}\in V},
\end{equation}
where Einstein summation convention is assumed. Note that the electric fields in the surface term
of \eqref{eq:SHPolarization} should be evaluated just beneath the particle boundary.

The surface nonlinear susceptibility tensor $\hat{\bm{\chi}}_s^{(2)}$ has 27 components, but for
homogeneous and isotropic surfaces only three of them are independent. These components are
$\chi_{s,\perp\perp\perp}^{(2)}$, $\chi_{s,\perp\parallel\parallel}^{(2)}$, and
$\chi_{s,\parallel\perp\parallel}^{(2)} = \chi_{s,\parallel\parallel\perp}^{(2)}$ \cite{Shen},
defined with respect to the local orthonormal coordinate system located at each point on the
surface, and the subscripts $\bot$ and $\parallel$ refer to the orientations perpendicular and
parallel to the boundary, respectively. The surface nonlinear polarization sheet,
$\bm{\mathcal{P}}_{s}^{\Omega}$, defined by the relation
$\mathbf{P}_s^{\Omega}(\mathbf{r})\equiv\bm{\mathcal{P}}_{s}^{\Omega}(\mathbf{r})\delta(\mathbf{r}-\mathbf{r}_{s})$,
can be expressed in spherical coordinates associated to the sphere as:
\begin{subequations}\label{eq:SurfPolcomp}
\begin{align}
& \mathcal{P}_{s,r}^{\Omega} = \epsilon_0\left[\hat{\chi}_{s,\perp\perp\perp}^{(2)}E_r^{\omega}E_r^{\omega} + \hat{\chi}_{s,\perp\parallel\parallel}^{(2)}(E_{\theta}^{\omega}E_{\theta}^{\omega} + E_{\phi}^{\omega}E_{\phi}^{\omega})\right]\label{eq:SurfPolcomp_a}, \\
& \mathcal{P}_{s,\theta}^{\Omega} =2\epsilon_0 \hat{\chi}_{s,\parallel\perp\parallel}^{(2)}E_r^{\omega}E_{\theta}^{\omega}    \label{eq:SurfPolcomp_b}, \\
& \mathcal{P}_{s,\phi}^{\Omega} =2\epsilon_0
\hat{\chi}_{s,\parallel\perp\parallel}^{(2)}E_r^{\omega}E_{\phi}^{\omega} \label{eq:SurfPolcomp_c}.
\end{align}
\end{subequations}

The bulk nonlinear polarization contribution $\vect{P}_b^{\Omega}$ can be cast in the following
form \cite{Bloembergen}:
\begin{equation}
\vect{P}_b^{\Omega} = \epsilon_0 \beta \vect{E}^{\omega} \nabla \cdot \vect{E}^{\omega}  +
\epsilon_0 \gamma \nabla (\vect{E}^{\omega} \cdot \vect{E}^{\omega}) + \epsilon_0 \delta'
(\vect{E}^{\omega} \cdot \nabla)\vect{E}^{\omega}, \label{eq:BulkPolar}
\end{equation}
where $\beta$, $\gamma$ and $\delta'$ are material parameters. The first term in
\eqref{eq:BulkPolar} vanishes in homogeneous materials since there is no net charge density present
in the bulk of the material, whereas the last term can be neglected as well, as most theoretical
models predict \cite{Sipe1}.

The electromagnetic fields at the SH frequency, $\Omega$, are generated by the nonlinear
polarization sources, $\vect{P}_n^{\Omega}$, $n=1,2,\ldots,N$, associated to each sphere. These
nonlinear fields satisfy the following \textit{inhomogeneous} system of equations:
\begin{subequations}\label{eq:MaxwinsSH}
\begin{align}
&\nabla \times \vect{H}_{n}^{\Omega}(\vect{r})+i\Omega
\epsilon_{n}^{\Omega}\vect{E}_{n}^{\Omega}(\vect{r})=-i\Omega\vect{P}_{b,n}^{\Omega}(\vect{r}),\quad
\vect{r}\in V_{n} \label{eq:MaxwinsSH_a}, \\
&\nabla \times \vect{E}_{n}^{\Omega}(\vect{r})-i\Omega\mu_{0}\vect{H}_{n}^{\Omega}(\vect{r})= 0.
\quad \vect{r}\in V_{n}, \label{eq:MaxwinsSH_b}
\end{align}
\end{subequations}
Moreover, the external nonlinear fields satisfy the following \textit{homogeneous} equations:
\begin{subequations}\label{eq:MaxwoutSH}
\begin{align}
&\nabla \times \vect{H}_{e}^{\Omega}(\vect{r}) + i\Omega \epsilon_{e}\vect{E}_{e}^{\Omega}(\vect{r})=0, \quad \vect{r}\in V_{e}, \label{eq:MaxwoutSH_a} \\
&\nabla \times \vect{E}_{e}^{\Omega}(\vect{r}) - i\Omega  \mu_{0}\vect{H}_{e}^{\Omega}(\vect{r})=
0, \quad \vect{r}\in V_{e}. \label{eq:MaxwoutSH_b}
\end{align}
\end{subequations}
In addition to \eqref{eq:MaxwinsSH} and \eqref{eq:MaxwoutSH}, the components of the internal and
external nonlinear fields tangent to the surface $S_n$ of the $n^{th}$ sphere, $n=1,2,\ldots,N$,
are subject to the nonlinear boundary conditions (see Appendix B) \cite{Heinz}:
\begin{subequations}\label{eq:TancondSH}
\begin{align}
&\hat{\vect{n}}_n \times[\vect{E}_{e}^{\Omega}(\vect{r})-\vect{E}_n^{\Omega}(\vect{r})]=-\bm{\mathcal{M}}_n^{\Omega}, \quad \vect{r}\in S_{n}, \label{eq:TancondSH_a} \\
&\hat{\vect{n}}_n
\times[\vect{H}_{e}^{\Omega}(\vect{r})-\vect{H}_n^{\Omega}(\vect{r})]=\bm{\mathcal{J}}_n^{\Omega},
\quad \vect{r}\in S_{n}. \label{eq:TancondSH_b}
\end{align}
\end{subequations}
In these equations, the magnetic and electric surface currents,
$\bm{\mathcal{M}}_n^{\Omega}(\theta_{n},\phi_{n})$ and
$\bm{\mathcal{J}}_n^{\Omega}(\theta_{n},\phi_{n})$, respectively, are determined by the normal and
tangent contributions of the surface nonlinear polarization sheet,
$\bm{\mathcal{P}}_{s}^{\Omega}(\theta_{n},\phi_{n})$, with all these surface vector functions
depending only on the coordinates $(\theta_{n},\phi_{n})$ describing the tangent plane at
$\mathbf{r}_{n}$:
\begin{subequations}\label{eq:NonlCurr}
\begin{align}
&\bm{\mathcal{J}}_n^{\Omega}(\vect{r}) =  i\Omega \hat{\vect{n}}_n \times \left[\hat{\vect{n}}_n \times \bm{\mathcal{P}}_{s,n}^{\Omega}(\vect{r})\right], \quad \vect{r}\in S_{n}, \label{eq:NonlCurr_a}\\
&\bm{\mathcal{M}}_n^{\Omega}(\vect{r}) =\frac{1}{\epsilon_0} \hat{\vect{n}}_n \times \nabla_{S}
\left[\hat{\vect{n}}_n \cdot \bm{\mathcal{P}}_{s,n}^{\Omega}(\vect{r})\right],\quad \vect{r}\in
S_{n}, \label{eq:NonlCurr_b}
\end{align}
\end{subequations}
where $\nabla_{S}$ acts in the tangent plane at $\vect{r}\in S_{n}$.

In order to solve the systems of partial differential equations, \eqref{eq:MaxwinsSH} and
\eqref{eq:MaxwoutSH}, together with the boundary conditions \eqref{eq:TancondSH}, one has to find
the general solution of both homogeneous systems and the particular solution satisfying the
inhomogeneous system of equations \eqref{eq:MaxwinsSH}.

The two homogeneous systems of equations describing the SH process can be written as:
\begin{subequations}\label{eq:MaxwinsSHhom}
\begin{align}
&\nabla \times \bar{\vect{H}}_{n}^{\Omega}(\vect{r}) + i\Omega \epsilon_{n}^{\Omega}\bar{\vect{E}}_{n}^{\Omega}(\vect{r}) = 0,\quad \vect{r}\in V_{n}, \label{eq:MaxwinsSHhom_a} \\
&\nabla \times \bar{\vect{E}}_{n}^{\Omega}(\vect{r}) - i\Omega
\mu_{0}\bar{\vect{H}}_{n}^{\Omega}(\vect{r})= 0, \quad \vect{r}\in V_{n}, \label{eq:MaxwinsSHhom_b}
\end{align}
\end{subequations}
and
\begin{subequations}\label{eq:MaxwoutSHhom}
\begin{align}
&\nabla \times \bar{\vect{H}}_{e}^{\Omega}(\vect{r}) +\Omega i\epsilon_{e} \bar{\vect{E}}_{e}^{\Omega}(\vect{r}) = 0, \quad \vect{r}\in V_{e}, \label{eq:MaxwoutSHhom_a} \\
&\nabla \times \bar{\vect{E}}_{e}^{\Omega}(\vect{r}) - \Omega i \mu_{0}
\bar{\vect{H}}_{e}^{\Omega}(\vect{r})= 0, \quad \vect{r}\in V_{e}, \label{eq:MaxwoutSHhom_b}
\end{align}
\end{subequations}
where $\left\{\bar{\vect{E}}_{n}^{\Omega}, \bar{\vect{H}}_{n}^{\Omega}\right\}$ and
$\left\{\bar{\vect{E}}_{e}^{\Omega}, \bar{\vect{H}}_{e}^{\Omega}\right\}$, are the SH
electromagnetic field solutions satisfying the homogeneous systems of equations
\eqref{eq:MaxwinsSHhom} and \eqref{eq:MaxwoutSHhom}, valid inside the $n^{th}$ particle and in the
exterior domain, respectively. The external homogeneous field solution
$\left\{\bar{\vect{E}}_{e}^{\Omega}, \bar{\vect{H}}_{e}^{\Omega}\right\}$ must obey the
Silver-M\"{u}ller radiation condition. The general solution of the two systems \eqref{eq:MaxwinsSH}
and \eqref{eq:MaxwoutSH} is then:
\begin{subequations}\label{eq:ComplsolSHint}
\begin{align}
&\vect{H}_{n}^{\Omega}(\mathbf{r}) = \bar{\vect{H}}_{n}^{\Omega}(\mathbf{r}) + \vect{H}_{n}^{\Omega,p}(\mathbf{r}), \quad \vect{r}\in V_{n}, \\
&\vect{E}_{n}^{\Omega}(\mathbf{r}) = \bar{\vect{E}}_{n}^{\Omega}(\mathbf{r}) +
\vect{E}_{n}^{\Omega,p}(\mathbf{r}), \quad \vect{r}\in V_{n},
\end{align}
\end{subequations}
and
\begin{subequations}\label{eq:ComplsolSHext}
\begin{align}
&\vect{H}_{e}^{\Omega}(\mathbf{r}) = \bar{\vect{H}}_{e}^{\Omega}(\mathbf{r}), \quad \vect{r}\in V_{e},\\
&\vect{E}_{e}^{\Omega}(\mathbf{r}) = \bar{\vect{E}}_{e}^{\Omega}(\mathbf{r}), \quad \vect{r}\in
V_{e},
\end{align}
\end{subequations}
where $\left(\vect{E}_{n}^{\Omega,p}, \vect{H}_{n}^{\Omega,p}\right)$ is the particular solution of
the inhomogeneous system \eqref{eq:MaxwinsSH}. Considering that $\nabla\times\nabla
F(\mathbf{r})=0$ for any scalar function $F(\mathbf{r})$, it can be seen that a particular solution
of the inhomogeneous system \eqref{eq:MaxwinsSH} is given by:
\begin{subequations}\label{eq:PartsolSH}
\begin{align}
&\vect{H}_{n}^{\Omega,p}(\mathbf{r}) = 0, \quad \vect{r}\in V_{n} \\
&\vect{E}_{n}^{\Omega,p}(\mathbf{r}) = -
\frac{1}{\epsilon_{n}^{\Omega}}\vect{P}_{b,n}^{\Omega}(\vect{r}) = -
\frac{\epsilon_0\gamma}{\epsilon_{n}^{\Omega}} \nabla \left[\vect{E}_{n}^{\omega}(\mathbf{r}) \cdot
\vect{E}_{n}^{\omega}(\mathbf{r})\right], \quad \vect{r}\in V_{n}. \end{align}
\end{subequations}

Exploiting the nonlinear boundary conditions \eqref{eq:TancondSH} and decomposition of the
nonlinear SH fields to homogeneous and particular components, as per \eqref{eq:ComplsolSHint} and
\eqref{eq:ComplsolSHext}, we can express the boundary conditions for the SH homogeneous field
solutions as:
\begin{subequations}\label{eq:TancondSHhom}
\begin{align}
&\hat{\vect{n}}_n \times[\bar{\vect{E}}_{e}^{\Omega}(\vect{r})-\bar{\vect{E}}_n^{\Omega}(\vect{r})]=-\bm{\mathcal{M}}_n^{\Omega}(\vect{r}) + \hat{\vect{n}}_n \times \vect{E}_{n}^{\Omega,p}(\vect{r}),\quad \vect{r}\in S_{n}, \label{eq:TancondSHhom_a} \\
&\hat{\vect{n}}_n
\times[\bar{\vect{H}}_{e}^{\Omega}(\vect{r})-\bar{\vect{H}}_n^{\Omega}(\vect{r})]=\bm{\mathcal{J}}_n^{\Omega}(\vect{r}),
\quad \vect{r}\in S_{n}. \label{eq:TancondSHhom_b}
\end{align}
\end{subequations}
We stress that the nonlinear surface polarization source $\vect{P}_{s,n}^{\Omega}$ does not
explicitly enter in the equations describing the SH process, namely expressions
\eqref{eq:MaxwinsSH} and \eqref{eq:MaxwoutSH}, but rather it contributes to the nonlinear fields at
the SH \textit{via} the nonlinear boundary conditions imposed at the surface of the particles, as
per \eqref{eq:TancondSH}.

It can be easily seen that the homogeneous systems of equations \eqref{eq:MaxwinsSHhom} and
\eqref{eq:MaxwoutSHhom} at the SH, together with the corresponding boundary conditions
\eqref{eq:TancondSHhom}, are similar to the system of equations \eqref{eq:Maxwins} and
\eqref{eq:Maxwout} at the FF and the accompanying boundary conditions \eqref{eq:Tancond} for the
linear fields. Therefore, in similar fashion to the series expansion in terms of VSWFs of the
electromagnetic fields at the FF introduced in the preceding subsection, we proceed by expanding
the SH external and internal electric fields $\bar{\vect{E}}_{e}^{\Omega}$ and
$\bar{\vect{E}}_{n}^{\Omega}$, respectively, as \cite{Forestiere1}:
\begin{subequations}\label{eq:E-VSWFsh}
\begin{align}\label{eq:ExternalVSWFsh}
&\bar{\vect{E}}_{e}^{\Omega}(\vect{r}) = E_0^2 \sum\limits_{n=1}^{N} \sum\limits_{\nu\geq1} \left[
b_{n,\nu}^{\Omega} \vect{M}_{\nu}^{(3)} (k_e^{\Omega} \mathbf{r}_n) + a_{n,\nu}^{\Omega}
\vect{N}_{\nu}^{(3)} (k_e^{\Omega}
\mathbf{r}_n)\right],\\
\label{eq:InsideVSWFsh} &\bar{\vect{E}}_{n}^{\Omega}(\vect{r}) = E_0^2 \sum\limits_{\nu\geq1}
\left[ c_{n,\nu}^{\Omega} \vect{M}_{\nu}^{(1)} (k_{n}^{\Omega} \mathbf{r}_n) + d_{n,\nu}^{\Omega}
\vect{N}_{\nu}^{(1)} (k_{n}^{\Omega} \mathbf{r}_n)\right],
\end{align}
\end{subequations}
where the $\sim$$E_0^2$ dependence of the nonlinear fields on the amplitude of the (linear)
excitation field has been explicitly introduced and $k_e^{\Omega}$ and $k_{n}^{\Omega}$ are the
wave numbers in the background medium and interior of the $n^{th}$ sphere, respectively, calculated
at the frequency $\Omega$.

Similarly to the fundamental frequency VSWF-expansions of the fields introduced in
\eqref{eq:ScatteredVSWF} and \eqref{eq:InsideVSWF}, the unknown coefficients
$\{b_{n,\nu}^{\Omega},a_{n,\nu}^{\Omega}\}$ and $\{c_{n,\nu}^{\Omega},d_{n,\nu}^{\Omega}\}$ are
used in the series expansion of the SH fields in terms of VSWFs. Moreover, in conjunction with
\eqref{eq:E-VSWFsh}, the series expansions in terms of VSWFs of the magnetic fields
$\bar{\vect{H}}_{n}^{\Omega}$ and $\bar{\vect{H}}_{e}^{\Omega}$ can be easily obtained from
\eqref{eq:MaxwinsSHhom_b} and \eqref{eq:MaxwoutSHhom_b}, respectively.

A key step in the calculation of the unknown expansion coefficients
$\{b_{n,\nu}^{\Omega},a_{n,\nu}^{\Omega}\}$ and $\{c_{n,\nu}^{\Omega},d_{n,\nu}^{\Omega}\}$ of the
SH fields is the expansion of the vector functions in the r.h.s. of the nonlinear homogeneous
boundary conditions \eqref{eq:TancondSHhom}, related to the $n^{th}$ particle, in series of vector
spherical harmonic (VSH) functions (for the definition of these functions and their relation to
VSWFs $\vect{M}_{\nu}^{(J)}$ and $\vect{N}_{\nu}^{(J)}$ see Appendix A). The three VSH functions
are mutually orthogonal: the function $\vect{Y}_{\nu}$ points in the radial direction, whereas the
the other two functions, $\vect{X}_{\nu}$ and $\vect{Z}_{\nu}$, lie in the plane tangent to the
sphere. Therefore, the (tangential) SH homogeneous boundary conditions \eqref{eq:TancondSHhom} can
be expanded in series of (transverse) VSH functions $\vect{X}_{\nu}(\theta_n,\phi_n)$ and
$\vect{Z}_{\nu}(\theta_n,\phi_n)$ \cite{Forestiere1}:
\begin{subequations}\label{eq:TancondSHhomExp}
\begin{align}
&\bm{\mathcal{J}}_n^{\Omega} = \hat{\vect{n}}_n \times \left[ \frac{i E_0^2}{Z_{e}} \sum\limits_{\nu\geq1} \left( u_{n,\nu}^{'\Omega} \vect{X}_{\nu} + v_{n,\nu}^{'\Omega} \vect{Z}_{\nu}\right)\right], \label{eq:TancondSHhomExp_a} \\
&-\bm{\mathcal{M}}_n^{\Omega} + \hat{\vect{n}}_n \times \vect{E}_{n}^{\Omega, p}\big|_{S_n} =
\hat{\vect{n}}_n \times \left[E_0^2 \sum\limits_{\nu\geq1}\left( u_{n,\nu}^{''\Omega}
\vect{Z}_{\nu} + v_{n,\nu}^{''\Omega} \vect{X}_{\nu} \right)\right], \label{eq:TancondSHhomExp_b}
\end{align}
\end{subequations}
where $Z_e=\sqrt{\sfrac{\mu_0}{\epsilon_e}}$ is the impedance of the background medium and
$\{u_{n,\nu}^{'\Omega},v_{n,\nu}^{'\Omega}\}$ and $\{u_{n,\nu}^{''\Omega},v_{n,\nu}^{''\Omega}\}$
are Fourier coefficients, which can be determined by calculating certain integrals over the surface
of the unit sphere, utilizing the orthogonality properties of spherical harmonics
\cite{Dadap1,Varshalovich}. The surface integrations can be done analytically and the results are
expressed in terms of Clebsch-Gordan series, given in \cite{Forestiere1}.

The SH \textit{T}-matrix corresponding to the $n^{th}$ sphere, with $n =1, \ldots, N$, which
relates the external field coefficients, $\{b_{n,\nu}^{\Omega},a_{n,\nu}^{\Omega}\}$, to the
coefficients used in the expansion of the SH sources, $\{u_{n,\nu}^{'\Omega},v_{n,\nu}^{'\Omega}\}$
and $\{u_{n,\nu}^{''\Omega}, v_{n,\nu}^{''\Omega}\}$, can be written as:
\begin{equation}\label{eq:TmatrixsingleSH}
\mathbf{f}_n^{\Omega}  = \mathbf{T}_n^{\Omega} \mathbf{g}_n^{\Omega}.
\end{equation}

Similarly to the FF \textit{T}-matrix \eqref{eq:Tmatrixsingle}, the SH \textit{T}-matrix of a
single spherical particle, $\mathbf{T}_n^{\Omega}$, is a diagonal matrix with the entries computed
by imposing the nonlinear homogeneous boundary conditions \eqref{eq:TancondSHhom} to the SH field
expansions \eqref{eq:E-VSWFsh}. The vector $\mathbf{g}_n^{\Omega}$ contains the SH source-field
expansion coefficients, $\mathbf{g}_n^{\Omega}=[v_{n,\nu}^{'\Omega}, u_{n,\nu}^{'\Omega},
v_{n,\nu}^{''\Omega}, u_{n,\nu}^{''\Omega}]^{\mathrm{T}}$, and is determined by the FF field,
whereas the vector $\mathbf{f}_n^{\Omega}$ holds the coefficients of the series expansion in VSWFs
of the external SH field, $\mathbf{f}_n^{\Omega}=[b_{n,\nu}^{'\Omega}, a_{n,\nu}^{'\Omega},
b_{n,\nu}^{''\Omega}, a_{n,\nu}^{''\Omega}]^{\mathrm{T}}$, where $b_{n,\nu}^{\Omega} =
b_{n,\nu}^{'\Omega} + b_{n,\nu}^{''\Omega}$ and $a_{n,\nu}^{\Omega} = a_{n,\nu}^{'\Omega} +
a_{n,\nu}^{''\Omega}$.

The extension to a multiparticle system is straightforward and analogous to the approach used in
the case of the derivation of the FF multiparticle scattering matrix \eqref{eq:Tmatrixcluster}. The
total SH electric field exciting the $n^{th}$ sphere is equal to the sum of the fields radiated by
all the other spheres, and is given by:
\begin{align}\label{eq:IncFielnthParSH}
    \vect{E}_{n,ex}^{\Omega}(\mathbf{r}) = E^2_0 \sum\limits_{\substack{k=1\\k \ne n}}^{N}\left[\vect{M}_{\nu}^{(3)}(k_e^\Omega
    \mathbf{r}_k),\vect{N}_{\nu}^{(3)}(k_e^\Omega \mathbf{r}_k)\right]\mathbf{f}_k^{\Omega}, \\ \quad
    n=1,\ldots,N.\nonumber
\end{align}
This field can be expanded in the local coordinates associated to the $n^{th}$ sphere employing the
translation-addition matrices $\bm{\alpha}_{n,k}^{\Omega}$, evaluated at the SH frequency $\Omega$,
as:
\begin{align}\label{eq:IncFielnthParSHloc}
    \vect{E}_{n,ex}^{\Omega}(\mathbf{r}_n) = E^2_0 \sum\limits_{\substack{k=1\\k \ne n}}^{N}\left[\vect{M}_{\nu}^{(1)}(k_e^\Omega
    \mathbf{r}_n),\vect{N}_{\nu}^{(1)}(k_e^\Omega \mathbf{r}_n)\right]\bm{\alpha}_{n,k}^{\Omega}\mathbf{f}_k^{\Omega}, \\ \quad
    n=1,\ldots,N.\nonumber
\end{align}

Similarly to the procedure used at the FF, one expands the total SH field exciting the $n^{th}$
sphere, $\vect{E}_{n,ex}^{\Omega}(\mathbf{r}_n)$, into a series of regular VSWFs, with the
corresponding expansion coefficients forming the vector $\vect{e}_n^{\Omega}$. The equation
\eqref{eq:IncFielnthParSHloc} can then be written solely in terms of SH field expansion
coefficients as follows:
\begin{equation}\label{eq:IncFielnthParOrigCoeffSH}
 \mathbf{e}_n^{\Omega} = \sum\limits_{\substack{k=1\\k
 \ne n}}^{N} \bm{\alpha}_{n,k}^{\Omega}\mathbf{f}_k^{\Omega}, \quad n=1,\ldots,N.
\end{equation}
In conjunction with the definition of the single sphere SH \textit{T}-matrix
\eqref{eq:TmatrixsingleSH}, a relation between the SH external field expansion coefficients can now
be derived:
\begin{equation}\label{eq:ExtFielnthParOrigCoeffSH}
    \mathbf{f}_n^{\Omega} =\sum\limits_{\substack{k=1\\k \ne n}}^{N} \mathbf{T}_n^\Omega
    \bm{\alpha}_{n,k}^{\Omega}\mathbf{f}_k^{\Omega}, \quad n=1,\ldots,N.
\end{equation}
Finally, this equation can be recast in a matrix form:
\begin{equation}\label{eq:TmatrixclusterSH}
\begin{bmatrix}
\textbf{I} & -\textbf{T}_{1}^{\Omega} \boldsymbol{\alpha}_{1,2}^{\Omega} & \cdots &  -\textbf{T}_{1}^{\Omega} \boldsymbol{\alpha}_{1,N}^{\Omega}\\
-\textbf{T}_{2}^{\Omega} \boldsymbol{\alpha}_{2,1}^{\Omega} & \textbf{I} & \cdots &  -\textbf{T}_{2}^{\Omega} \boldsymbol{\alpha}_{2,N}^{\Omega} \\
\vdots & \vdots & \ddots & \vdots \\
-\textbf{T}_{N}^{\Omega} \boldsymbol{\alpha}_{N,1}^{\Omega} &
-\textbf{T}_{N}^{\Omega} \boldsymbol{\alpha}_{N,2}^{\Omega} & \cdots & \textbf{I}
\end{bmatrix}
\begin{bmatrix}
\textbf{f}_{1}^{\Omega} \\
\textbf{f}_{2}^{\Omega} \\
\vdots \\
\textbf{f}_{N}^{\Omega}
\end{bmatrix}
=
\begin{bmatrix}
\textbf{T}_{1}^{\Omega} \textbf{g}_{1}^{\Omega} \\
\textbf{T}_{2}^{\Omega} \textbf{g}_{2}^{\Omega} \\
\vdots \\
\textbf{T}_{N}^{\Omega} \textbf{g}_{N}^{\Omega} \\
\end{bmatrix},
\end{equation}
or written more compactly,
\begin{equation}
\mathbf{S}^{\Omega} \mathbf{F}^{\Omega}  = \mathbf{G}^{\Omega}. \label{eq:ScattmatclustSH}
\end{equation}

The matrix $\mathbf{S}^{\Omega}$ is the SH multiparticle scattering matrix and $\mathbf{F}^{\Omega}
= \{\mathbf{f}_n^{\Omega}\}$ and $\mathbf{G}^{\Omega} =
\{\textbf{T}_{n}^{\Omega}\mathbf{g}_n^{\Omega}\} $ represent the vector of the unknown SH external
field expansion coefficients and the vector containing the expansion coefficients of the SH
excitation field, respectively.

By solving the linear system \eqref{eq:ScattmatclustSH} one obtains for each spherical scatterer
the expansion coefficients of the SH external field. Using these coefficients, the external and
internal fields at the SH, given by \eqref{eq:E-VSWFsh}, as well as the SH SCS can be obtained in a
similar fashion to the procedure used in the FF case. The calculation of the internal SH fields
\eqref{eq:ComplsolSHint} and the SH ACS is somewhat more convoluted because it requires the
evaluation of the particular solution \eqref{eq:PartsolSH} determined by the bulk polarization,
$\mathbf{P}_{b}^{\Omega}$. This requires an accurate series expansion in VSWFs of the particular SH
electric field solution $\vect{E}_{n}^{\Omega,p}$, $n=1,\ldots,N$. This is achieved using the
relations for the products of VSHs given in \cite{Varshalovich}. Once the particular solution is
computed, the total internal SH fields can be obtained using \eqref{eq:ComplsolSHint}, and
consequently the SH ACS can be calculated as \cite{Biris}:
\begin{equation}\label{eq:AbscrosssectSH}
\sigma_{abs}^{\Omega} = \frac{P_{abs}^{\Omega}}{P_{inc}},
\end{equation}
where $P_{inc}$ is the power of the incident plane wave at the FF and $P_{abs}^{\Omega}$ is the
total dissipated power in all scatterers at the SH frequency:
\begin{equation}\label{eq:TotalPowerSH}
P_{abs}^{\Omega} = \frac{1}{2} \sum_{n=1}^{N}\iiint\limits_{V_n}
\mathfrak{Re}\left[\sigma_n^{\Omega} \vect{E}_n^{\Omega}(\vect{r})
\vect{E}_n^{\Omega*}(\vect{r})\right] d\mathbf{r},\quad \vect{r}\in V_{n},
\end{equation}
where $\sigma_n^{\Omega}$ is the conductivity of the $n^{th}$ sphere at the SH frequency.

\section{Numerical examples}\label{Examples}
In this section, using several generic examples, we illustrate how our numerical method can be used
to compute the electromagnetic fields and the scattering and absorption cross sections, both at the
FF and SH frequency. In order to validate our numerical method, we compare in several cases our
results with those computed using CST Studio \cite{CSTStudio}, a commercial software based on the
FEM. We consider nanospheres made of gold (one of the most used noble metals in plasmonics) or
silicon (a dielectric material widely used in nonlinear nanooptics) embedded in vacuum. The system
is illuminated by a plane wave propagating along a direction characterized by the angles
$\theta_{inc} = \pi/4$ and $\phi_{inc} = \pi/2$ and polarized along the $\hat{\bm{\theta}}$
direction, the amplitude of the electric field being $E_0=\SI{1}{\volt\per\meter}$.

\begin{figure}[!t]
    \centering
    \includegraphics[width=\columnwidth]{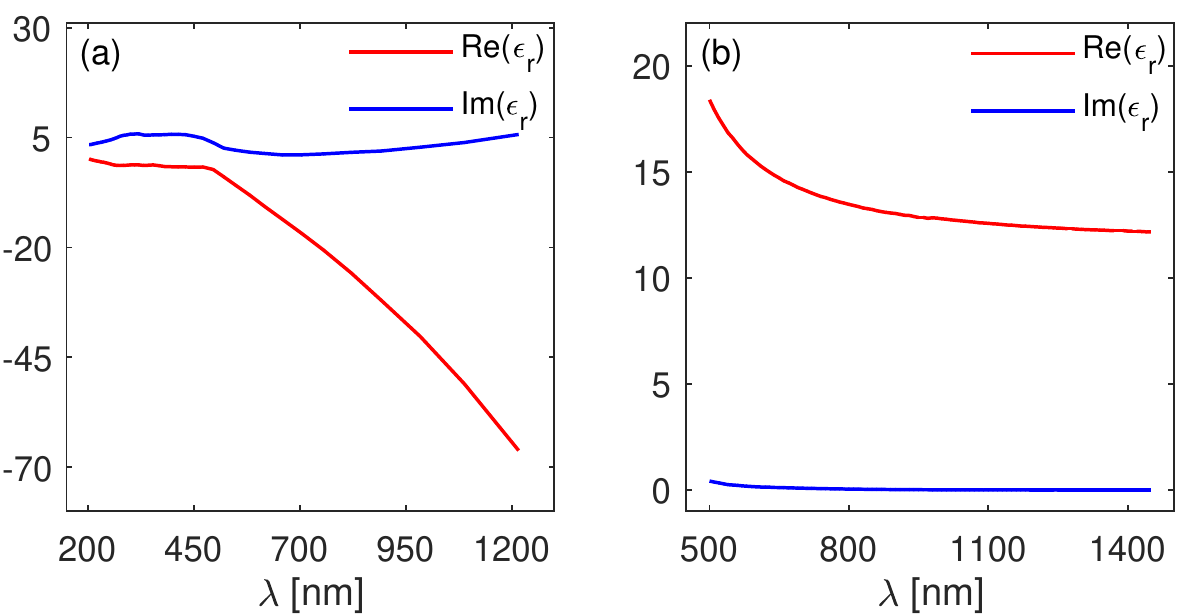}
\caption{(a) Real and imaginary parts of the relative permittivity of gold. (b) The same as in (a),
but corresponding to silicon.} \label{fig:Epsilons}
\end{figure}

The permittivities of gold and silicon are computed by interpolating the experimental data provided
in Johnson and Christy's \cite{Johnson} and Schinke's \cite{Schinke}, respectively, see
Fig.~\ref{fig:Epsilons}. Moreover, according to the hydrodynamic model \cite{Sipe1}, the components
of third-rank surface and bulk susceptibility tensors of gold are given by:
\begin{subequations}\label{eq:RS}
\begin{align}
&\chi_{s,\perp\perp\perp}^{(2)} = -\frac{a}{4} \left[\epsilon_{r}(\omega) - 1\right]
\frac{\epsilon_0
e}{m\omega^2},\label{eq:ksiperperper}\\
&\chi_{s,\perp\parallel\parallel}^{(2)} = -\frac{b}{2} \left[\epsilon_{r}(\omega) -1\right]
\frac{\epsilon_0
e}{m\omega^2}, \label{eq:ksiparparper} \\
&\gamma  = -\frac{d}{8} \left[\epsilon_{r}(\omega) - 1\right] \frac{\epsilon_0 e}{m\omega^2},
\label{eq:gammaBul}
\end{align}
\end{subequations}
and $\chi_{s,\parallel\perp\parallel}^{(2)} = \chi_{s,\parallel\parallel\perp}^{(2)}=0$
\cite{Corvi,Guyot}. The coefficients $a$, $b$, and $d$ in \eqref{eq:RS} are the so-called
Rudnick-Stern parameters \cite{Rudnick}, and in the case of the hydrodynamic model they are $a=1$,
$b=-1$, and $d=1$ \cite{Sipe1}. The nonlinear second-order susceptibilities of silicon are assumed
to be frequency independent over the spectral range of interest. Their values are:
$\chi_{s,\perp\perp\perp}^{(2)} =\SI{65e-19}{\meter\squared\per\volt}$,
$\chi_{s,\perp\parallel\parallel}^{(2)} = \SI{3.5e-19}{\meter\squared\per\volt}$, and $\gamma =
\SI{1.3e-19}{\meter\squared\per\volt}$ \cite{Falasconi}.

\subsection{Convergence of the numerical method}
We begin the analysis of our numerical method with a study of its convergence characteristics. To
this end, it should be noted that there are two main physical quantities that our numerical method
is most suited to compute, namely scattering and absorption cross-sections, which are globally
defined quantities, and the optical near-field, which is a local physical quantity. Therefore,
analyzing the convergence of the calculations of these physical quantities will provide important
information about both the global and local convergence properties of our numerical method.
\begin{figure}[!b]
\centering
\includegraphics[width=\columnwidth]{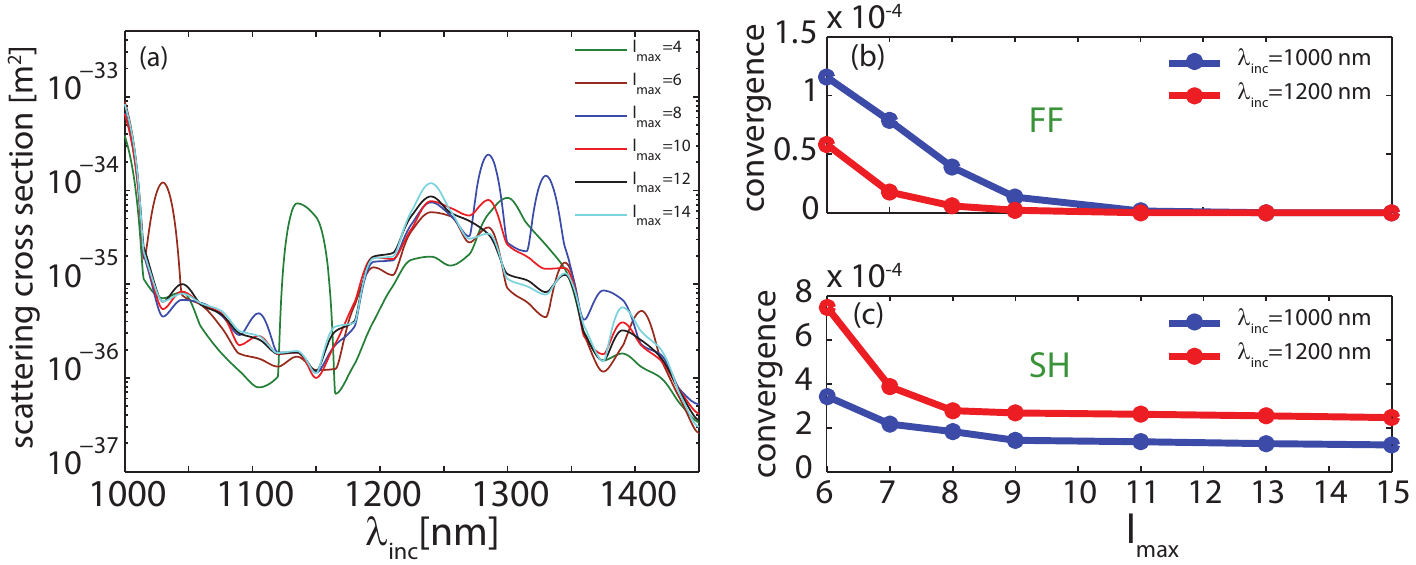}
\caption{(a) Scattering cross-section calculated at the SH for a system of two silicon nanospheres
with radius $R_1=R_2=\SI{300}{\nano\meter}$ and centers at $O_1(0,0,0)$ and
$O_2(0,0,\SI{800}{\nano\meter})$. (b) and (c) The error of the near-field,
$\mathcal{E}_{l_{max}}^{\omega,\Omega}$, defined in \eqref{eq:norm}, determined at the FF and SH frequency,
respectively.} \label{fig:Conv}
\end{figure}

The system we used in testing the convergence of the method consists of two silicon nanospheres
with radius $R_1=R_2=\SI{300}{\nano\meter}$ and centers at $O_1(0,0,0)$ and
$O_2(0,0,\SI{800}{\nano\meter})$. For this system, we calculated the SCS spectrum at the FF and SH,
as well as the FF and SH electric fields in a grid of $M=2500$ points located in the $xy$-plane at
$z=\SI{400}{\nano\meter}$, and covering the domain
$[-400,400]\times[-400,400]~\si{\square\nano\meter}$. As the convergence properties of the linear
TMM have been extensively studied, we focus here on the cross-sections at the SH (see
Fig.~\ref{fig:Conv}a). Furthermore, the calculations were performed for increasing values of the
cut-off of the orbital index, $l_{max}$, from $l_{max}=4$ up to $l_{max}=17$. In order to
characterize the convergence of the near-field, we computed the norm
\begin{align}\label{eq:norm}
\mathcal{E}_{l_{max}}^{\omega,\Omega} = \frac{1}{M} \sqrt{\sum_{i=1}^{M}\left|
\mathbf{E}_{l_{max}}^{\omega,\Omega}(\mathbf{r}_{i})-\mathbf{E}_{l_{max}=17}^{\omega,\Omega}(\mathbf{r}_{i})\right|^{2}},
\end{align}
and plotted the results in Fig.~\ref{fig:Conv}b (FF) and Fig.~\ref{fig:Conv}c (SH).

These results demonstrate that the calculation of both the SCS at the SH and the linear and
nonlinear near-fields converge asymptotically as $l_{max}$ increases (note that the SCS at the FF
converges much faster), which suggests both global and local convergence of the method. This is
particularly important as it appears that the numerical method converges rather fast even for
nanoparticles with relatively large index of refraction ($n_{Si}\simeq3.45$), namely in a case when
the near-field is strongly inhomogeneous.

\subsection{Scattering from a single nanosphere}
\begin{figure}[!b]
\centering
\includegraphics[width=\columnwidth]{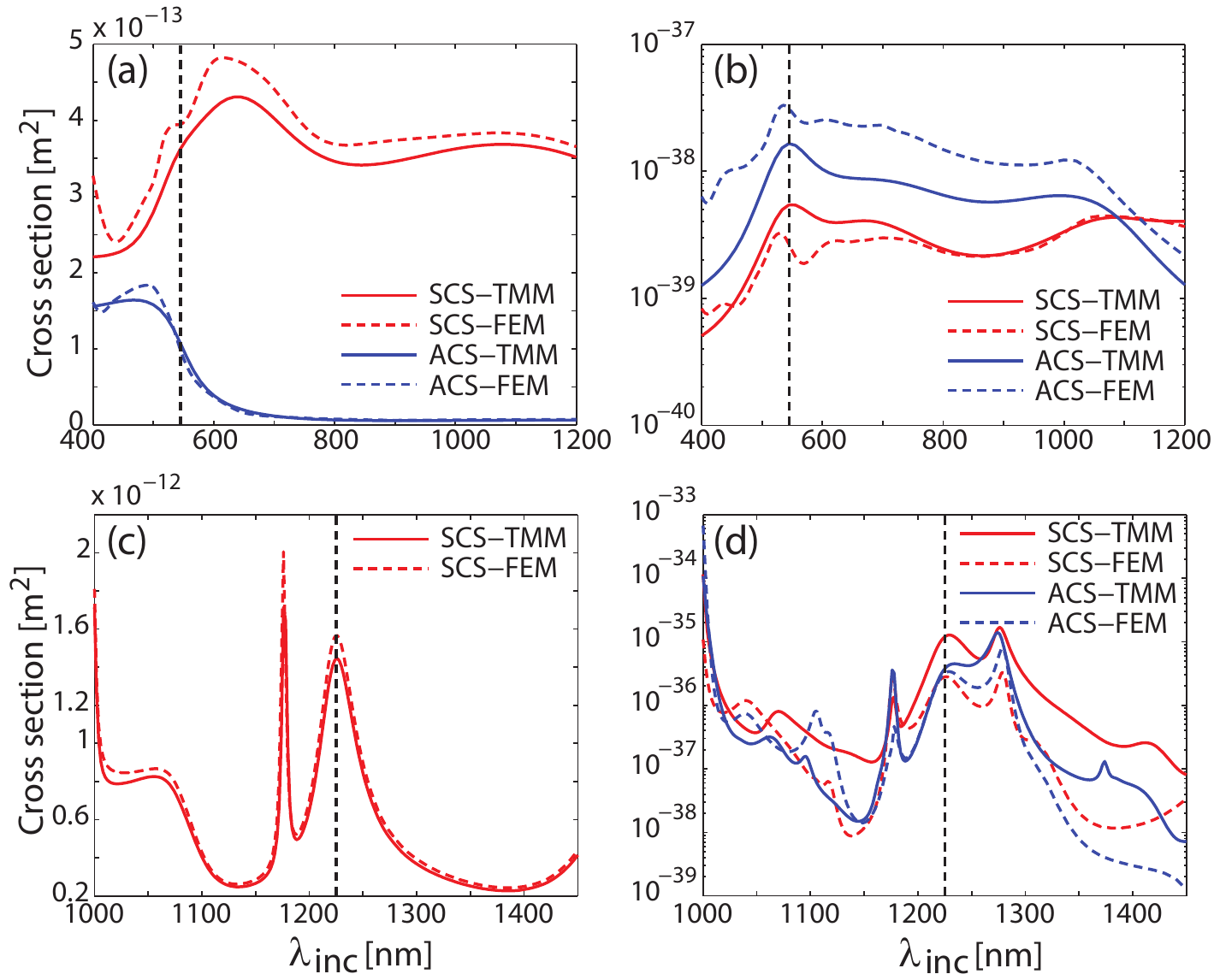}
\caption{Scattering and absorption cross-sections calculated at the FF (left panels) and SH (right
panels). The top and bottom panels correspond to a gold nanosphere with radius
$R=\SI{200}{\nano\meter}$ and a silicon nanosphere with radius $R=\SI{300}{\nano\meter}$,
respectively. The vertical lines indicate the input wavelengths at which the spatial profiles of
the near-field are calculated.} \label{fig:Qsca_1AuSi}
\end{figure}

In the first example, we analyzed the linear and nonlinear scattering from a single gold nanosphere
with radius $R=\SI{200}{\nano\meter}$ and a silicon nanosphere with radius
$R=\SI{300}{\nano\meter}$, the calculated linear and nonlinear SCSs and ACSs being presented in
Fig.~\ref{fig:Qsca_1AuSi}. These cross-sections have been calculated both with the TMM and the FEM,
so that we could further validate our numerical method. As it can be seen from
Fig.~\ref{fig:Qsca_1AuSi}, there is a good agreement between the spectra of the SCSs and ACSs, both
at the FF and SH, calculated using our numerical method based on the TMM and the commercial FEM
software.
\begin{figure}[!b]
\centering
\includegraphics[width=\columnwidth]{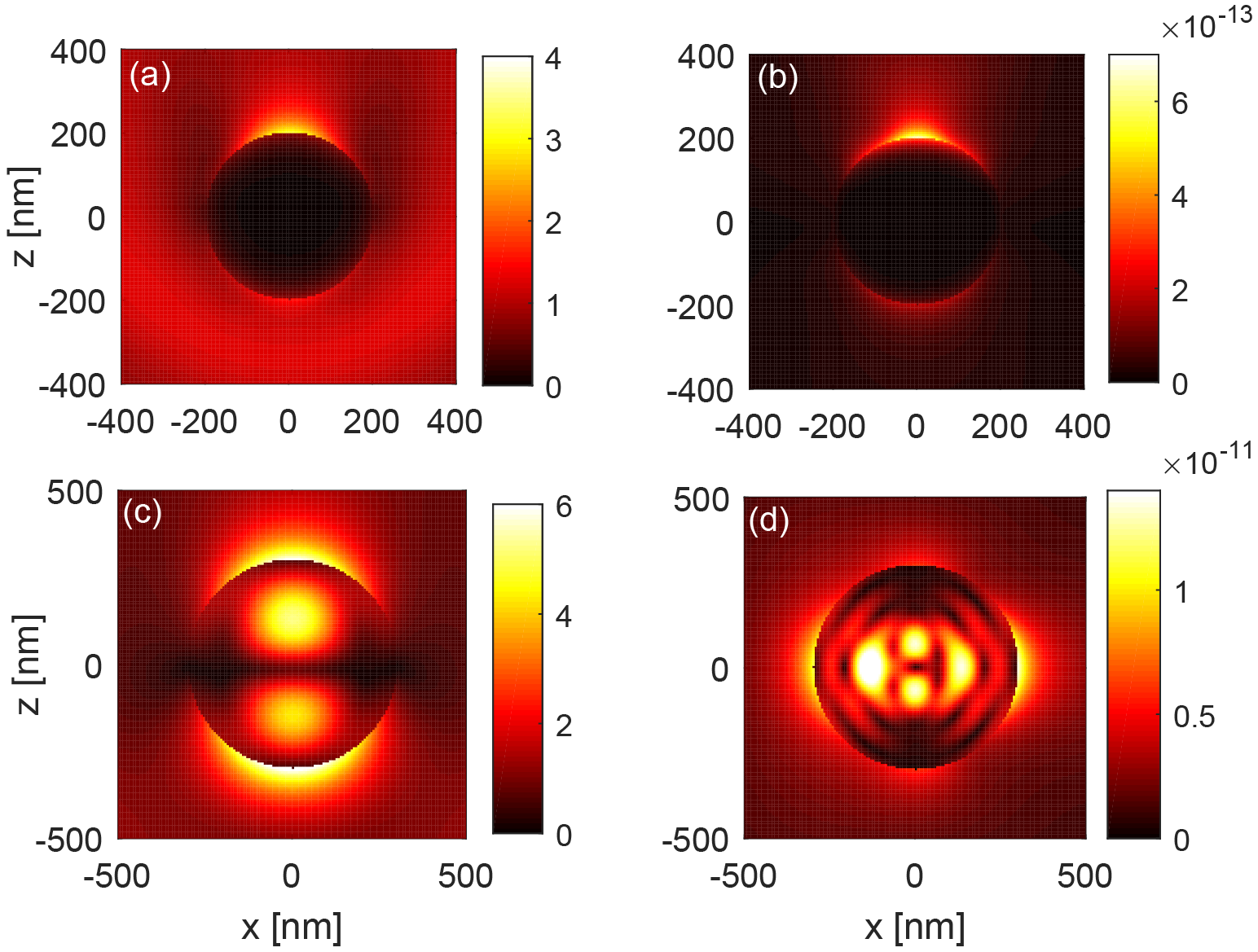}
\caption{Spatial distribution of the electric field amplitude at the FF (left panels) and SH (right
panels), calculated for a single gold nanosphere (top panels) at the incident wavelength
$\lambda_{inc} = \SI{545}{\nano\meter}$ and for a single silicon nanosphere (bottom panels) at the
incident wavelength $\lambda_{inc} = \SI{1225}{\nano\meter}$. The fields are determined in the
$xz$-plane.} \label{fig:Eabs_1AuSi}
\end{figure}

In the case of a single sphere, the TMM results are analytical, since the elements of the transfer
matrix are equal to the Mie coefficients, which can be calculated analytically. Therefore, the
discrepancy between the TMM and FEM results, especially notable in the SH regime, as per Figs.
\ref{fig:Qsca_1AuSi}b and \ref{fig:Qsca_1AuSi}d, are perhaps due to the inaccuracies introduced by
the discretization intrinsic to the FEM. More specifically, the electromagnetic field at the SH is
strongly inhomogeneous and confined near the surface of the nanosphere, where the nonlinear
polarization sources are located. This makes it difficult to be accurately resolved with finite
elements. Moreover, approximately 25000 tetrahedral finite elements were used for a single
frequency calculation both in the fundamental and second-harmonic case for gold and silicon
spheres. By contrast, in our approach, only 390 VSWSs were needed, resulting in a significant
memory reduction as compared to the FEM.

The spectra of the SCS of the gold nanosphere, at the FF and SH, show a pronounced maximum at the
incident wavelengths $\lambda_{inc} = \SI{640}{\nano\meter}$ and $\lambda_{inc} =
\SI{545}{\nano\meter}$, respectively. These spectral peaks are due to the excitation of localized
surface plasmons on the metallic nanosphere, and are characterized by a large enhancement of the
local optical field. In the case of the silicon sphere, the SCS spectrum at the FF has two peaks,
which in this case correspond to electric dipole and magnetic dipole Mie resonances.

In Fig.~\ref{fig:Eabs_1AuSi} we present the spatial distribution of the amplitude of the electric
field, $\lvert\vect{E}\rvert$, determined in the \textit{xz}-plane, at both the FF (shown in
Figs.~\ref{fig:Eabs_1AuSi}a and \ref{fig:Eabs_1AuSi}c) and SH (shown in Figs.~\ref{fig:Eabs_1AuSi}b
and \ref{fig:Eabs_1AuSi}d). In the case of the gold sphere, the field profile is plotted at the
incident wavelength $\lambda_{inc} = \SI{545}{\nano\meter}$ corresponding to the maximum in the SH
scattering spectrum, whereas the field profile for the silicon sphere is computed at the incident
wavelength $\lambda_{inc} = \SI{1225}{\nano\meter}$ and reveals the electric dipole character of
the first resonance in the FF scattering spectrum. Moreover, it can be seen that the optical field
does not penetrate into the gold sphere irrespective of the wavelength. In addition,
unsurprisingly, the field at the SH is more inhomogeneous than at the FF, for both gold and silicon
nanospheres.
\begin{figure}[!b]
\centering
\includegraphics[width=\columnwidth]{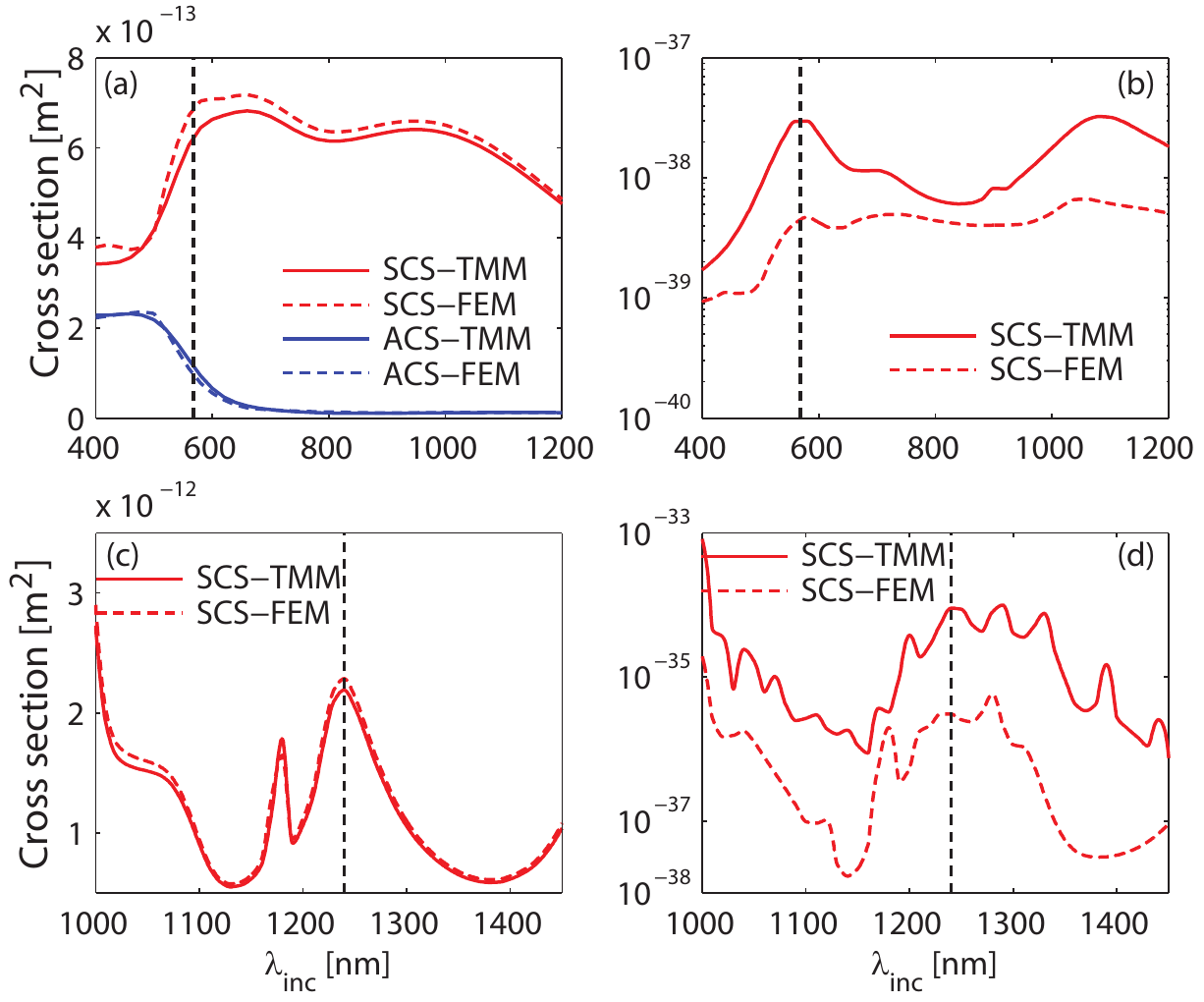}
\caption{Scattering and absorption cross-section spectra at the FF (left panels) and SH (right
panels). The top and bottom panels correspond to a gold nanodimer composed of two nanospheres with
radii $R_1=\SI{150}{\nano\meter}$ and $R_2=\SI{200}{\nano\meter}$ centered in $O_1(0,0,0)$ and
$O_2(0,0,\SI{550}{\nano\meter})$, and for a silicon nanodimer composed of two identical spheres
with radii $R_1=R_2=\SI{300}{\nano\meter}$ centered in $O_1(0,0,0)$ and
$O_2(0,0,\SI{800}{\nano\meter})$, respectively. The vertical lines indicate the input wavelengths
at which the spatial profiles of the near-field are calculated.} \label{fig:Qsca_2AuSi}
\end{figure}

\subsection{Second-harmonic generation in nanodimers}
Since in the case of a single sphere the translation-addition matrices are not employed, this is
not a particularly challenging test for the full capabilities of the TMM. Therefore, in the second
example, we analyze nanodimers made of gold and silicon nanospheres. In the case of the gold
nanodimer, the spheres are centered at $O_1(0,0,0)$ and $O_2(0,0,\SI{550}{\nano\meter})$, the
corresponding radii being $R_1=\SI{150}{\nano\meter}$ and $R_2=\SI{200}{\nano\meter}$. The silicon
nanodimer is composed of two identical spheres with radii $R_1=R_2=\SI{300}{\nano\meter}$, located
at $O_1(0,0,0)$ and $O_2(0,0,\SI{800}{\nano\meter})$. Finite-element method calculations have also
been employed, using about 42000 tetrahedral elements for the gold dimer and 30000 elements for the
silicon one. On the other hand, we used merely 780 VSWFs for the silicon and gold dimers. These
examples are relevant for practical applications since strong near-fields can be generated in the
space in-between the spheres, thus allowing for the design of highly sensitive sensing devices.
\begin{figure}[!b]
\centering
\includegraphics[width=\columnwidth]{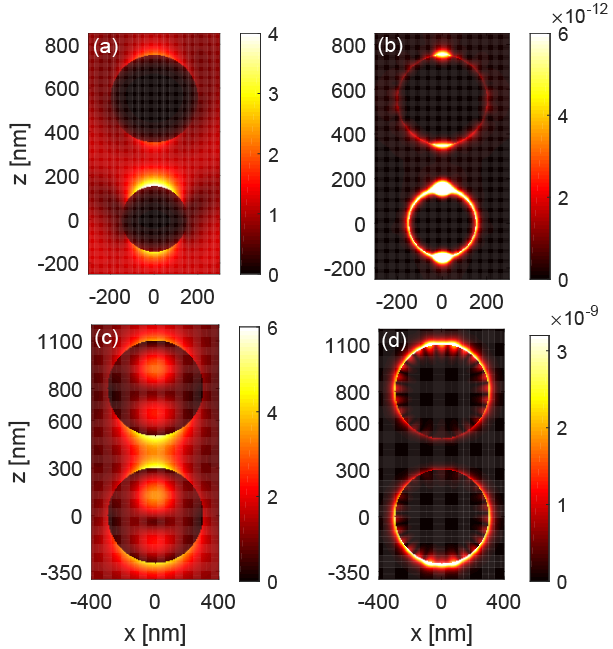}
\caption{Spatial distribution of the electric field amplitude at the FF (left panels) and SH (right
panels), calculated for the gold nanodimer (top panels) at the incident wavelength $\lambda_{inc} =
\SI{568}{\nano\meter}$ and for the silicon nanodimer (bottom panels) at the incident wavelength
$\lambda_{inc} = \SI{1240}{\nano\meter}$. The fields are determined in the $xz$-plane.}
\label{fig:Eabs_2AuSi}
\end{figure}

The spectrum of the SCS at the FF corresponding to the gold dimer shows a maximum at $\lambda_{inc}
= \SI{660}{\nano\meter}$, whereas two peaks are observed in the spectrum of the SCS at the SH, at
the incident wavelengths $\lambda_{inc} = \SI{560}{\nano\meter}$ and $\lambda_{inc} =
\SI{1080}{\nano\meter}$. The SCS spectra of the silicon nanodimer have a more complex structure. In
particular, in the linear regime, two spectral peaks are observed for incident wavelengths of
similar values as those in the case of a single silicon nanosphere, but slightly redshifted due to
inter-particle interactions (compare Fig.~\ref{fig:Qsca_1AuSi}c and Fig.~\ref{fig:Qsca_2AuSi}c). A
similar situation is seen in the case of the SCS spectra at the SH. A series of spectral peaks can
be observed, and again they are somewhat redshifted as compared to the case of a single silicon
sphere (compare Fig.~\ref{fig:Qsca_1AuSi}d and Fig.~\ref{fig:Qsca_2AuSi}d). These spectral peaks
are of plasmonic nature in the case of the gold dimer and originate from interacting Mie resonances
in the case of the silicon dimer.

A perceptive reader has perhaps noticed that we did not present the nonlinear ACS spectra for the
dimer cases. The main reason for this is that we found that they converge very slowly. This is
explained by the fact that the ACS depends on the near-field, which is a physical quantity that
converges very slowly, chiefly due to the inhomogeneous nature of the local field. To be more
specific, the internal field expansion coefficients at the SH, $c_{n,\nu}^{\Omega}$ and
$d_{n,\nu}^{\Omega}$, vary very slowly with the cut-off value of the orbital index, $l_{max}$, and
therefore a large number of VSWFs must be included in the series expansion in order for the
near-field (and consequently ACS) to converge. Moreover, we have noticed that when $l_{max}$ is
very large, the entries of translation-addition matrices can become numerically unstable. Numerical
errors introduced in the calculation of SH scattering field expansion coefficients are amplified
when they are used to obtain internal SH field coefficients. This propagation of numerical errors
particularly affects the SH absorption spectra of multiple nanospheres, and more so when their
radii are smaller than the incident wavelength. Nevertheless, the physical quantity of practical
interest is the SCS, as the power absorbed by clusters of particles is difficult to measure
experimentally.

The spatial profile of the amplitude of the electric field computed in the \textit{xz}-plane is
shown in Fig.~\ref{fig:Eabs_2AuSi}. In the case of the gold nanodimer, we choose the incident
wavelength $\lambda_{inc} = \SI{568}{\nano\meter}$, namely at the location of the second peak in
the SH spectrum of the SCS. In the silicon dimer case, we choose the wavelength $\lambda_{inc} =
\SI{1240}{\nano\meter}$ to correspond to one of the Mie resonance wavelength. Interestingly enough,
the field distribution corresponding to this latter case suggests that the resonance is a
whispering-gallery mode \cite{Biris2}.
\begin{figure}[!b]
\centering
\includegraphics[width=\columnwidth]{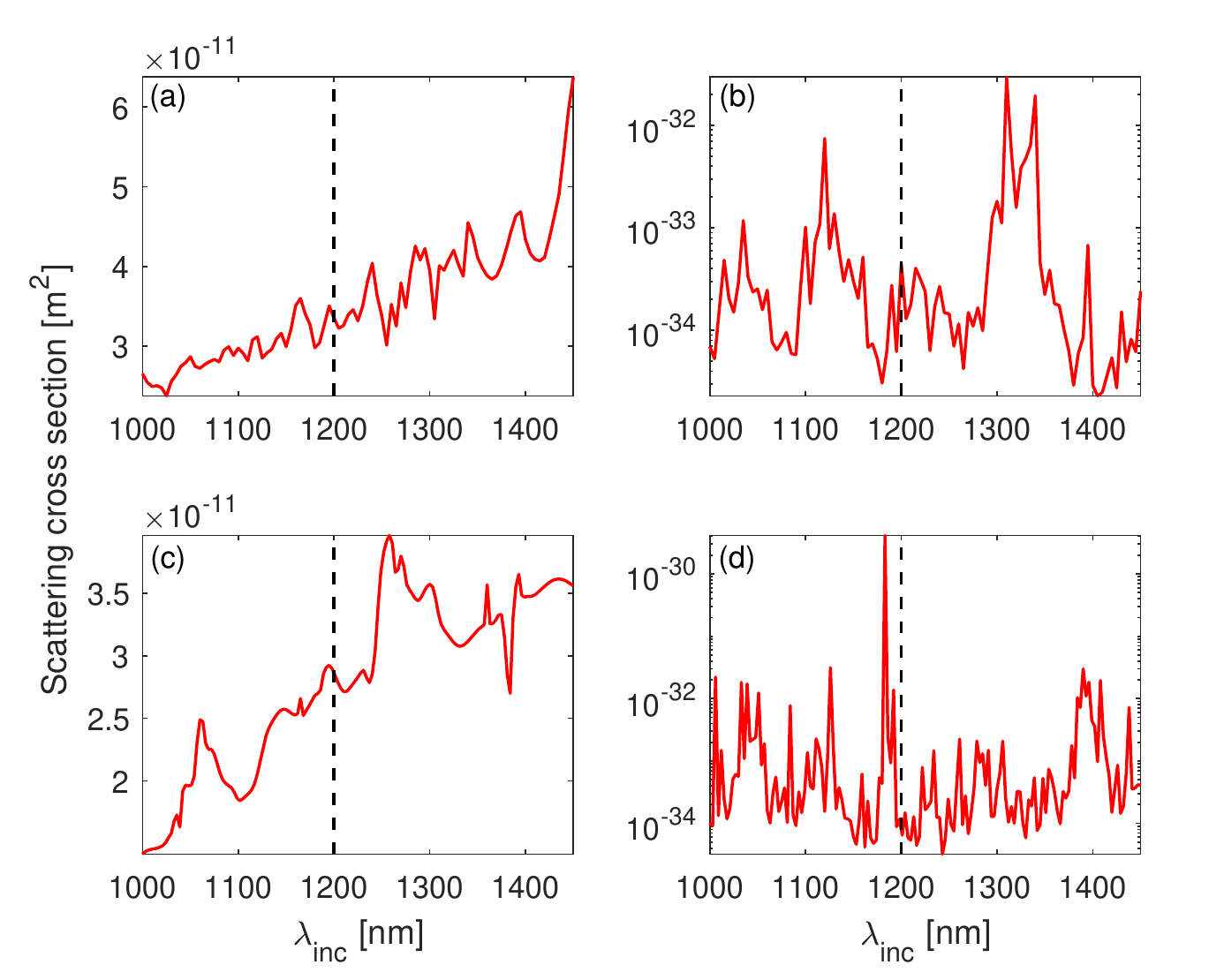}
\caption{Scattering cross-section spectra at the FF (left panels) and SH (right panels). The top
and bottom panels correspond to the centrosymmetric and non-centrosymmetric lattices, respectively.
Spectra of absorption cross-sections at the FF are not plotted since the losses are negligible in
this frequency range. The vertical line indicates the wavelength at which the field profile is
calculated.} \label{fig:Qsca_cubarr}
\end{figure}

\subsection{Second-harmonic generation in finite cubic lattices of nanospheres}
In this example, we consider nanospheres arranged in two cubic lattices. We assume that the
nanospheres in the two cubic lattices are made of silicon (a centrosymmetric material) and are
located in the first octant of the Cartesian coordinate system, aligned with the $x$-, $y$- and
$z$-axes. The first lattice contains $5\times5\times5=125$ identical nanospheres arranged so that
the cluster is centrosymmetric. The second one is a non-centrosymmetric cubic array consisting of 8
zincblende unit cells containing a total of 95 spheres. This kind of crystal structure is found,
for example, in GaAs or BAs, compounds widely used in the semiconductor industry. In the case of
the centrosymmetric cluster, the nanospheres have the same radius, $R_n=\SI{400}{\nano\meter}$,
$n=1,\ldots,125$, and are equally spaced with the center-to-center distance,
$d=\SI{850}{\nano\meter}$, thus forming a cubic array with side length equal to
$\SI{3.4}{\micro\meter}$. The non-centrosymmetric lattice contains two different sets of spheres:
the first set contains larger spheres, with radius $R_n=\SI{420}{\nano\meter}$, $n=1,\ldots,63$,
located at the vertices and side midpoints of the zincblende unit cells, whereas the other spheres
have radius $r_n=\SI{250}{\nano\meter}$, $n=1,\ldots,32$ and are located inside the unit cells. The
side length of this cluster is $\SI{4}{\micro\meter}$. We do not present here results of FEM
calculations because for this problem the memory requirements are prohibitive. By contrast, in our
TMM calculations we used 336 VSWFs per nanosphere, resulting in a total of 42000 basis functions
for the centrosymmetric lattice and 31920 for the non-centrosymmetric one.
\begin{figure}[!t]
\centering
\includegraphics[width=\columnwidth]{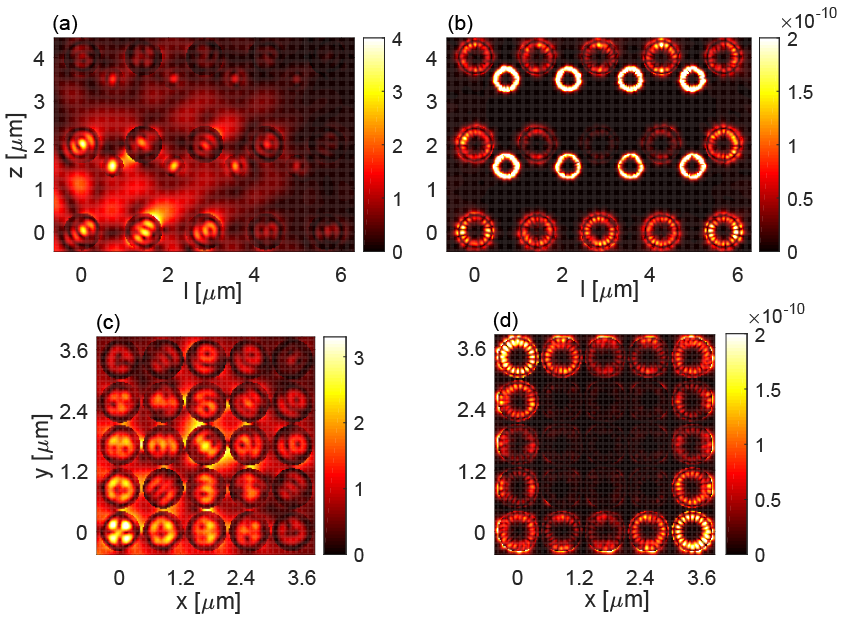}
\caption{Distribution of the electric field amplitude at FF (left panels) and SH (right panels),
calculated at $\lambda_{inc} = \SI{1200}{\nano\meter}$ for a non-centrosymmetric (top panels) and
centrosymmetric (bottom panels) cluster of silicon nanospheres arranged in a cubic lattice. Top
(bottom) panels correspond to the plane defined by the normal
$(\hat{\mathbf{i}}-\hat{\mathbf{j}})/\sqrt{2}$ (the $xy$-plane bisecting the cluster through its
center).} \label{fig:Eabs_cubarr}
\end{figure}

In Fig.~\ref{fig:Qsca_cubarr} we plot the SCS spectra, at the fundamental and second-harmonic
frequencies, for both types of lattices. It can be seen that they present an irregular aspect,
which reflects an intricate interplay between the resonances of different spheres and their optical
coupling. Moreover, the optical near-fields are presented in Fig.~\ref{fig:Eabs_cubarr}. Thus, in
the upper panels of this figure, the fields are presented in the diagonal plane of the cluster
defined by the normal $(\hat{\mathbf{i}}-\hat{\mathbf{j}})/\sqrt{2}$, whereas in the lower panels
of this same figure, they are calculated in the $xy$-plane bisecting the structure through its
center. These field profiles illustrate an important phenomenon related to the connection between
the SHG properties and the symmetry characteristics of the cluster of spheres. Thus, as it can be
observed in Figs.~\ref{fig:Eabs_cubarr}a and \ref{fig:Eabs_cubarr}b, the optical field penetrates
inside the non-centrosymmetric cluster both at the FF and SH frequency. By contrast, according to
the lower panels of Fig.~\ref{fig:Eabs_cubarr}, in the case of the centrosymmetric cluster the
fields are present inside the structure at FF only, whereas at the SH they are expelled from the
cluster and are mainly concentrated at its outer boundary (see Fig.~\ref{fig:Eabs_cubarr}d). The
reason for this is that the cluster in Figs.~\ref{fig:Eabs_cubarr}c and \ref{fig:Eabs_cubarr}d is
centrosymmetric and as such the (local) ``bulk'' second-order effective susceptibility cancels. The
``surface'' second-order effective susceptibility, however, has a nonzero value and therefore SH is
generated at the surface of the cluster. By contrast, the structure formed from zincblende unit
cells is non-centrosymmetric and thus the SHG is allowed in the bulk of the system, too.

\subsection{Second-harmonic generation in an irregular distribution of nanospheres}
\begin{figure}[!t]
\centering
\includegraphics[width=\columnwidth]{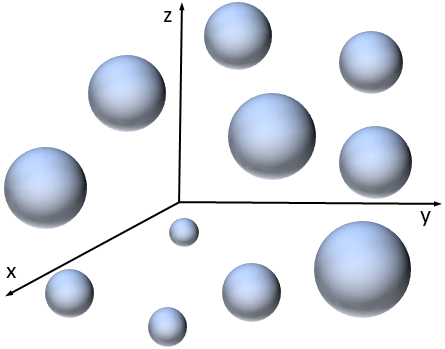}
\caption{Schematics of the cluster of silicon nanoparticles, drawn at scale, considered in the last
numerical example.} \label{fig:ElSpheres}
\end{figure}

The last example considered is the scattering from an irregular distribution of 11 silicon
nanoparticles. This is a good test for the versatility of our algorithm, since there are no
constraints imposed on the size and position of nanospheres. The only condition required to be
fulfilled is that the spheres do not touch (see Fig.~\ref{fig:ElSpheres}). In this example, we
chose the nanospheres radii to be in the interval
$R_{n}\in[\SI{80}{\nano\meter},\SI{310}{\nano\meter}], n = 1, \ldots, 11$.

\begin{figure}[!t]
\centering
\includegraphics[width=\columnwidth]{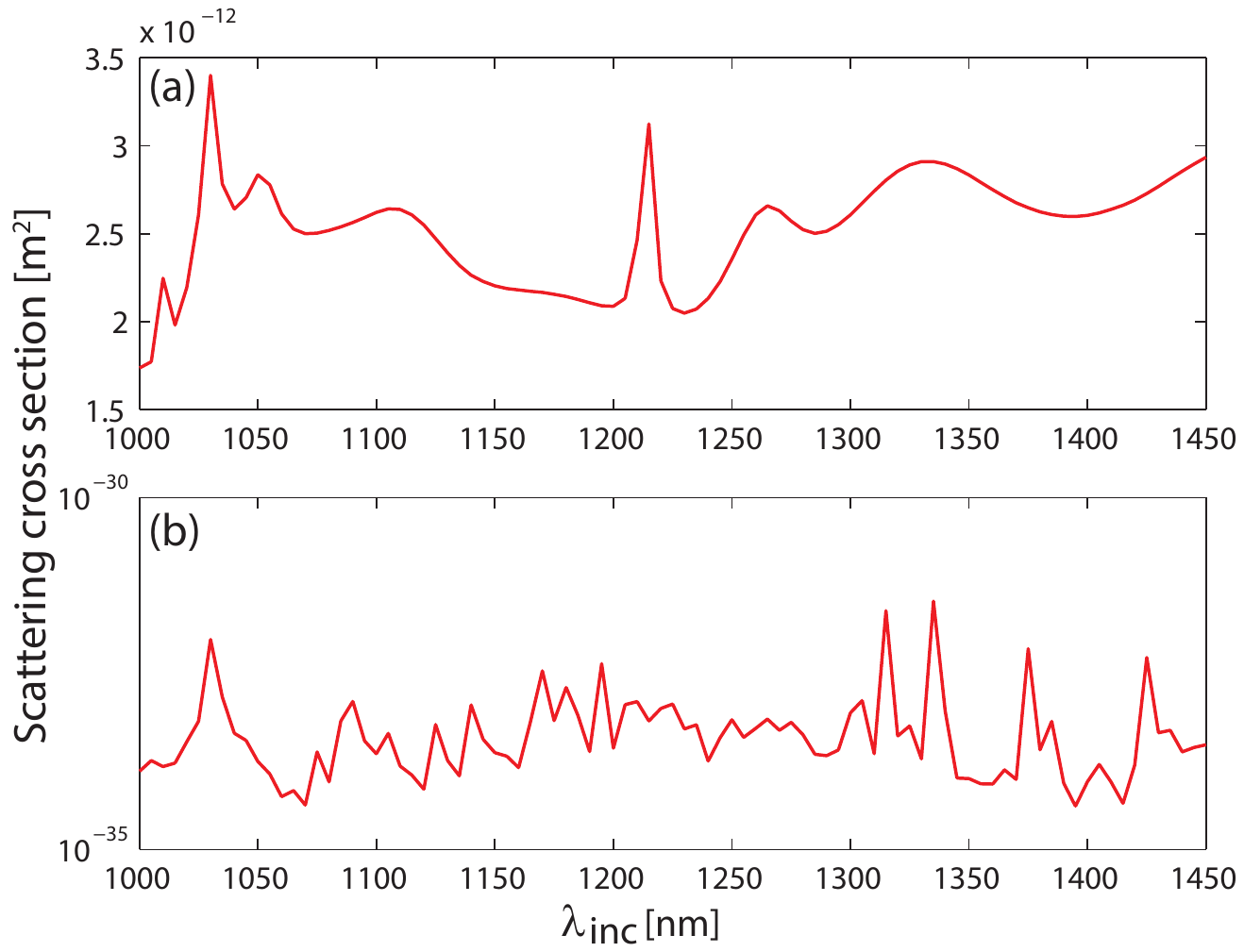}
\caption{Spectra of scattering cross-section at the FF (a) and SH (b), computed for an irregular
cluster of 11 silicon nanospheres. Spectra of absorption cross-section at the FF is not plotted
since the losses are negligible in this frequency range.} \label{fig:Qsca_11Si}
\end{figure}

In Fig.~\ref{fig:Qsca_11Si} we show the SCS spectra of the ensemble of silicon nanospheres. Two
peaks are revealed in the spectrum at the FF, one at $\lambda_{inc} = \SI{1030}{\nano\meter}$ and
one at $\lambda_{inc} = \SI{1215}{\nano\meter}$ (as per Fig.~\ref{fig:Qsca_11Si}a), whereas the SH
SCS spectrum presents a much larger number of resonances (see Fig.~\ref{fig:Qsca_11Si}b).

\section{Conclusion}\label{CONCL}
In this article we have introduced a \textit{T}-matrix method for fast and accurate electromagnetic
wave scattering analysis of nanosphere clusters. Our numerical method can be used to characterize
linear and nonlinear (second-harmonic) optical responses of the particle system. Different from
previous studies, we include both surface and bulk contributions to the nonlinear polarization
source. Following the classical \textit{T}-matrix approach, the fields oscillating at fundamental
frequency are expanded in a suitable set of vector spherical harmonics and the  transfer matrix of
the cluster is obtained by taking into account multiple wave scattering among the particles and
linear boundary conditions on the boundaries of the nanospheres. After the fundamental field
expansion coefficients are obtained, they are used in the computation of second-harmonic
polarization sources. In the second step of our method, the SH \textit{T}-matrix is formed by
incorporating the mutual coupling between spheres at second-harmonic and nonlinear boundary
conditions on their boundary-surfaces.

The proposed technique appears to have improved performance and lower memory imprint when compared
to standard finite element method, while maintaining the overall accuracy in the scattering
spectra. Furthermore, the algorithm lends itself to efficient parallelization on multi-core
high-performance computer architectures since the submatrices representing the electromagnetic
coupling between particles can be computed independently. This fact, along with state-of-the-art
parallel iterative solvers, pave the way for fast and accurate numerical analysis of clusters
containing thousands of nanoparticles.

\section*{Declaration of Competing Interest}
The authors declare that they have no known competing financial interests or personal relationships
that could have appeared to influence the work reported in this article.

\section*{CRediT authorship contribution statement}
Ivan Sekulic: Methodology, Software, Investigation, Writing - original draft, review \& editing.
Jian Wei You: Methodology, Investigation, Writing - review \& editing. Nicolae C. Panoiu:
Conceptualization, Methodology, Writing - original draft, review \& editing, Supervision.

\section*{Funding}
This work was supported by the European Research Council (ERC) (Grant no. ERC-2014-CoG-648328).

\section*{Acknowledgement}
We acknowledge the use of the UCL Legion High Performance Computing Facility (Legion@UCL) and
associated support services in the completion of this work.

\section*{Appendix A}\label{Maths}
In this Appendix we introduce the definition and main properties of vector spherical harmonics
(VSHs) used in our study to represent the electromagnetic fields, as well as to expand the plane
waves.
\subsection*{Definition of vector spherical wave functions}
Given the unit vector $\hat{\vect{r}} = \vect{r}/r$, where $\vect{r}$ is the position vector of a
point, $P$, we define three sets of mutually orthogonal VSHs: longitudinal VSHs, $\vect{Y}_{lm}$,
which are oriented along $\hat{\vect{r}}$, and two sets of transverse VSHs, $\vect{X}_{lm}$ and
$\vect{Z}_{lm}$, oriented perpendicular to $\hat{\vect{r}}$ \cite{Varshalovich,Stout2}. The triplet
$\{\vect{X}_{lm},\vect{Y}_{lm},\vect{Z}_{lm}\}$ of VSHs form a right-handed system of vectors and
are defined as:
\begin{subequations}
\begin{align}
&\vect{Y}_{lm}(\theta,\phi) =  Y_{lm}(\theta,\phi)\hat{\vect{r}},\label{eq:Xm1}\\
&\vect{X}_{lm}(\theta,\phi) = \frac{1}{\sqrt{l(l+1)}} \hat{\vect{r}} \times \nabla_{\hat{\vect{r}}}
Y_{lm}(\theta,\phi),\label{eq:Xzero}\\
&\vect{Z}_{lm}(\theta,\phi) = \frac{1}{\sqrt{l(l+1)}} \nabla_{\hat{\vect{r}}}
Y_{lm}(\theta,\phi).\label{eq:Xp1}
\end{align}
\label{eq:X}
\end{subequations}

Furthermore, we define the regular VSWFs $\vect{M}_{lm}^{(1)}$ and $\vect{N}_{lm}^{(1)}$ as
follows:
\begin{subequations}\label{eq:regularVSWF}
\begin{align}
&\vect{M}_{lm}^{(1)}(kr,\theta,\phi) = -j_l(kr) \vect{X}_{lm}, \\
&\vect{N}_{lm}^{(1)}(kr,\theta,\phi) = \frac{1}{k} \nabla \times \vect{M}_{lm}^{(1)}, \\
&\vect{M}_{lm}^{(1)}(kr,\theta,\phi) = \frac{1}{k} \nabla \times \vect{N}_{lm}^{(1)},
\end{align}
\end{subequations}
where $k$ is the wave number calculated at the particular frequency in the medium of choice and
$j_l(kr)$ are spherical Bessel functions of the first kind and order $l$. The radiative (outgoing)
VSWFs $\vect{M}_{lm}^{(3)}$ and $\vect{N}_{lm}^{(3)}$ are defined by substituting the spherical
Bessel functions in \eqref{eq:regularVSWF} with (outgoing) spherical Hankel functions,
$h_l^{(1)}(kr)$, of the first kind and order $l$.

Scalar spherical harmonics can be defined in terms of associated Legendre functions $P_l^m$ as
\cite{Stout2}:
\begin{equation}
Y_{lm}(\theta, \phi) = \gamma_{lm} \sqrt{l(l+1)} P_l^m(\cos \theta) e^{im\phi},
\label{eq:ScaSphHar}
\end{equation}
where $\gamma_{lm}$ is the normalization coefficient given by:
\begin{equation}
\gamma_{lm} = \sqrt{\frac{(2l + 1)(l - m)!}{4 \pi l(l + 1)(l + m)!}}. \label{eq:Normcoeff}
\end{equation}

In these definitions the scalar spherical harmonic functions $Y_{lm}(\hat{\mathbf{r}})$ and the
three kinds of VSHs are normalized to unity over a sphere:
\begin{subequations}\label{eq:normSphHarm}
\begin{align}
&\iint_{\Omega_{\hat{\mathbf{r}}}}Y_{lm}^{*}(\hat{\mathbf{r}})Y_{l^{\prime}m^{\prime}}(\hat{\mathbf{r}})d\Omega_{\hat{\mathbf{r}}}
= \delta_{ll^{\prime}}\delta_{mm^{\prime}},\\
&\iint_{\Omega_{\hat{\mathbf{r}}}}\mathbf{A}_{lm}^{*}(\hat{\mathbf{r}})\cdot
\mathbf{B}_{l^{\prime}m^{\prime}}(\hat{\mathbf{r}})d\Omega_{\hat{\mathbf{r}}} =
\delta_{ll^{\prime}}\delta_{mm^{\prime}}\delta_{AB},
\end{align}
\end{subequations}
where $\mathbf{A}$ and $\mathbf{B}$ can be any of the functions $\mathbf{X}$, $\mathbf{Y}$, and
$\mathbf{Z}$.

Finally, the normalized VSHs can be defined in terms of associated Legendre functions as:
\begin{subequations}\label{eq:normLegd}
\begin{align}
&\vect{Y}_{lm}(\theta,\phi) = \gamma_{lm}\sqrt{l(l+1)} P_l^m (\cos \theta)
e^{im\phi}\hat{\vect{r}}, \label{eq:Xm1Legd}\\
&\vect{X}_{lm}(\theta,\phi) = \gamma_{lm} \left[-\frac{im}{\sin \theta} P_l^m (\cos
\theta)e^{im\phi} \hat{\boldsymbol{\theta}} +\frac{d}{d \theta} P_l^m (\cos \theta)e^{im\phi}
\hat{\boldsymbol{\phi}} \right], \label{eq:XzeroLegd}\\
&\vect{Z}_{lm}(\theta, \phi) = \gamma_{lm} \left[ \frac{d}{d \theta} P_l^m (\cos \theta)e^{im\phi}
\hat{\boldsymbol{\theta}} + \frac{im}{\sin \theta} P_l^m (\cos \theta)e^{im\phi}
\hat{\boldsymbol{\phi}} \right]. \label{eq:Xp1Legd}
\end{align}
\end{subequations}

\subsection*{Expansion of plane waves in series of vector spherical wave functions}
The series expansion of a plane wave in terms of WSVFs is given in \eqref{eq:IncidentVSWF}, with
the expansion coefficients $q_{lm}^{\omega}$ and $p_{lm}^{\omega}$ calculated as follows
\cite{Stout2}:
\begin{subequations}\label{eq:PLWexpCoeff}
\begin{align}
q_{lm}^{\omega} &= -4\pi i^l \vect{X}_{lm}^{*}(\hat{\vect{k}}_{inc}) \cdot \hat{\vect{e}}_{inc},\\
p_{lm}^{\omega} &= -4\pi i^{l+1} \vect{Z}_{lm}^{*}(\hat{\vect{k}}_{inc}) \cdot
\hat{\vect{e}}_{inc},
\end{align}
\end{subequations}
where the symbol ``$*$'' denotes complex conjugation, $\hat{\vect{k}}_{inc}$ is the unit vector in
the direction of the wave propagation, and $\hat{\vect{e}}_{inc}$ is the polarization direction of
the plane wave.

\section*{Appendix B}\label{NlBC}
In this Appendix, we introduce the boundary conditions for the electromagnetic fields, valid in the
presence of a nonlinear surface polarization sheet, $\bm{\mathcal{P}}_s^{\Omega}(\mathbf{r})$, with
support on a surface $\mathcal{S}$ separating two optical media. The interface is characterized by
the normal, $\hat{\mathbf{n}}$, which we assume that points from medium $1$ to medium $2$. Then,
the fields satisfy the following boundary conditions \cite{Heinz}:
\begin{subequations}\label{eq:TancondSHtan}
\begin{align}
    &\Delta\mathbf{E}_{\parallel} = -  \frac{1}{\epsilon^{\prime}} \nabla_{S}
    \left[\hat{\vect{n}} \cdot \bm{\mathcal{P}}_s^{\Omega}(\mathbf{r})\right], \quad \vect{r}\in\mathcal{S}, \label{eq:TancondSHtan_a}\\
    &\Delta\mathbf{D}_{\perp} = -\nabla_{S} \cdot \bm{\mathcal{P}}_s^{\Omega}(\mathbf{r}), \quad \vect{r}\in\mathcal{S}, \label{eq:Normcond_a}\\
    &\Delta\mathbf{H}_{\parallel}= i\Omega \hat{\vect{n}} \times \bm{\mathcal{P}}_s^{\Omega}(\mathbf{r}), \quad
    \vect{r}\in\mathcal{S}, \label{eq:TancondSHtan_b}\\
    &\Delta\mathbf{B}_{\perp} = 0, \quad \vect{r}\in\mathcal{S}, \label{eq:Normcond_b}
\end{align}
\end{subequations}
where $\Omega$ is the frequency, the symbol $\parallel$ ($\perp$) refers to the field component
parallel (perpendicular) to the interface, $\nabla_{S}$ is the restriction of the operator $\nabla$
to the tangent plane, and, for a vector field $\mathbf{V}$, we defined $\Delta
\mathbf{V}=\mathbf{V}_{2}-\mathbf{V}_{1}$, where $\mathbf{V}_{1}$ ($\mathbf{V}_{2}$) is the value
of the field in medium $1$ ($2$) at a location infinitesimally close to the interface. Note that in
deriving \eqref{eq:TancondSHtan}, one assumes that the fields depend on time as $e^{-i \Omega t}$.
Using the fact that for a vector $\mathbf{V}$ the components tangent and perpendicular onto
$\mathcal{S}$ are $\mathbf{V}_{\parallel}=-\hat{\vect{n}}\times\hat{\vect{n}}\times\mathbf{V}$ and
$\mathbf{V}_{\perp}=(\hat{\vect{n}}\cdot\mathbf{V})\hat{\vect{n}}$, respectively, one can easily
show that \eqref{eq:TancondSHtan_a} and \eqref{eq:TancondSHtan_b} with
$\epsilon^{\prime}=\epsilon_{0}$ are equivalent to the boundary conditions \eqref{eq:TancondSH}.

%Manual citation list

%\bibliography{Notes}
%\bibliographystyle{aipauth4-1}

\end{document}